\input mtexsis
\input epsf.sty
\def\M{{\cal M}}
\def\cont{{\,{\underline{\;\;\,}}{\kern -0.23em}|\;}}
\def\4{{\scriptscriptstyle 0}}

\def\x{{\bf x}}

\def\boh{{\rm I{\kern -0.20em}H{\kern -0.20em}I}}
\def\bom{{\rm I{\kern -0.20em}M{\kern -0.20em}I}}
\def\real{{\rm I{\kern -0.20em}R}}
\def\comps{{\rm C{\mkern -8.5mu}
           {{\vrule height 6.8pt width 0.65pt depth -0.63pt}}}{\mkern 8.5mu}}

\def\proj{{\rm I{\kern -0.20em}P}}
\def\ident{{1{\kern -0.36em}1}}

\def\subo{{}_{{}_{\scriptscriptstyle {\rm O}}}}

\def\suba{{}_{{}_{\scriptscriptstyle 1}}}
\def\subb{{}_{{}_{\scriptscriptstyle 2}}}
\def\subc{{}_{{}_{\scriptscriptstyle 3}}}
\def\semidirect{{\ooalign{\hfil\raise.07ex\hbox{s}\hfil\crcr\mathhexbox20D}}}
\def\longlongrightarrow{\relbar\joinrel\longrightarrow}

\def\tbf{\twelvepoint\bf}
\def\boxit#1{\vbox{\hrule\hbox{\vrule\kern3pt\vbox{\kern-13pt#1\kern-13pt}
              \kern3pt\vrule}\hrule}}

\null
\titlepage
\hoffset=-0.35cm
\headline={\hfil}
\footline={\hfil}
\title
{\bf{Why ~the ~Quantum ~Must ~Yield ~to ~Gravity}}
\endtitle
\vskip 0.25in
\center
Joy ~Christian
\smallskip\tenpoint
{\it Wolfson College, University of Oxford, Oxford OX2 6UD, United Kingdom}
\smallskip\tenpoint
{\it Electronic address} : {\it joy.christian@wolfson.oxford.ac.uk}
\endcenter
\vskip 0.30in
\baselineskip 0.5cm

\bigskip
\bigskip
\bigskip
\bigskip

\abstract
{\tenpoint After providing an extensive overview of the conceptual elements
-- such as Einstein's `hole argument' -- that underpin Penrose's proposal for
gravitationally induced quantum state reduction, the proposal is constructively
criticised. Penrose has suggested a mechanism for objective reduction of
quantum states with postulated collapse time
${\tau = \hbar/\Delta E\,}$, where ${\Delta E}$ is an ill-definedness in the
gravitational self-energy stemming from the profound conflict between the
principles of superposition and general covariance. Here it is argued that,
even if Penrose's overall conceptual scheme for the breakdown of quantum
mechanics is {\it unreservedly} accepted, his formula for the collapse time
of superpositions reduces to ${\tau\rightarrow \infty}$
(${\Delta E\rightarrow 0}$) in the strictly
Newtonian regime, which is the domain of his proposed
experiment to corroborate the effect. A suggestion is made to rectify this
situation. In particular, recognising the cogency of Penrose's
reasoning in the domain of full `quantum gravity', it is demonstrated that an
appropriate experiment which could in principle corroborate his argued
`macroscopic' breakdown of superpositions is {\it not} the
one involving non-rotating mass distributions as he has
suggested, but a Leggett-type SQUID or BEC experiment involving
superposed mass distributions in {\it relative rotation}. The
demonstration thereby brings out one of the distinctive characteristics of
Penrose's scheme, rendering it empirically distinguishable from
other state reduction theories involving gravity. As an aside, a new
geometrical measure of gravity-induced deviation from quantum mechanics
{\it \`a la} Penrose is proposed, but now for the canonical commutation
relation ${[Q,\,P]=i\hbar}$.}
\endabstract

\bigskip
\bigskip
\bigskip
\bigskip
\bigskip
\bigskip
\bigskip
\bigskip

\center
To appear in ~{\it Physics Meets Philosophy at the Planck Scale},
edited by C.~Callender and N.~Huggett (Cambridge University Press)
\endcenter
\endtitlepage

\hoffset=-0.35cm
\voffset=1.00cm
\superrefsfalse
\footline={\tenrm\hfil\folio\hfil}
\headline={\hfil}
\pageno=2
\tenpoint

\parskip 0.0cm
\parindent 0.00cm

\section{Introduction -- From Schr\"odinger's Cat to Penrose's `OR':}

Quantum mechanics -- one of our most fundamental and successful theories --
is infested with a range of deep philosophical difficulties collectively known
as the measurement
problem (Schr\"odinger 1935, Shimony 1963, Wheeler and Zurek 1983, Bell 1990).
In a nutshell the problem may be
stated as follows: If the orthodox formulation of quantum theory -- which in
general allows attributions of only objectively indefinite properties or
{\it potentialities} (Heisenberg 1958)
to physical objects -- is interpreted in compliance with
what is usually referred to as scientific realism, then one is faced with an
irreconcilable incompatibility between the nonnegotiable linearity of quantum
dynamics -- which governs evolution of the network of potentialities --
and the apparent definite or actual properties of the physical objects of our
`macroscopic' world. Moreover, to date no epistemic explanation
of these potentialities (e.g., in terms of `hidden variables')
has been completely successful (Shimony 1989). Thus, on the one hand there is
overwhelming experimental evidence in favour of the quantum mechanical
potentialities, supporting the view that they comprise a novel (i.e.,
classically uncharted) metaphysical modality of Nature situated between logical
possibility and actuality (Shimony 1978, 1993b, 140-162 and 310-322), and
on the other hand there is phenomenologically compelling proliferation of
actualities in our everyday world, including even in the {\it micro}biological
domain. The problem then
is that a universally agreeable mechanism for {\it transition}
between these two ontologically very different modalities -- i.e., transition
from the multiplicity of potentialities to various specific actualities -- is
completely missing. As delineated, this is clearly a very serious
{\it physical} problem. What is more, as exemplified by Shimony (1993a, 56),
the lack of a clear understanding of this apparent transition in the
world is also quite a `dark cloud' for any reasonable program of scientific
realism.

\parindent 0.75cm
\parskip 0.25cm

Not surprisingly, there exists a vast number of proposed solutions to the
measurement problem in the literature (Christian 1996), some of which
-- the Copenhagen interpretation (Bohr 1935) for example --
being almost congenital to quantum mechanics. Among these proposed
solutions there exists a somewhat dissident yet respectable tradition of ideas
-- going all the way back to Feynman's pioneering thoughts on the subject as
early as in mid-fifties (Feynman 1957) -- on a possible gravitational
resolution of the problem. The basic tenet of these proposals can hardly be
better motivated than in Feynman's own words. In
his Lectures on Gravitation (Feynman 1995, 12-13)
he devotes a whole section to the
issue, entitled ``On the philosophical problems in quantizing macroscopic
objects'', and contemplates on a possible breakdown of quantum mechanics:

\midfigure{monolith}
\hrule
\vskip 0.85cm
\centerline{\epsfysize=10cm \epsfbox{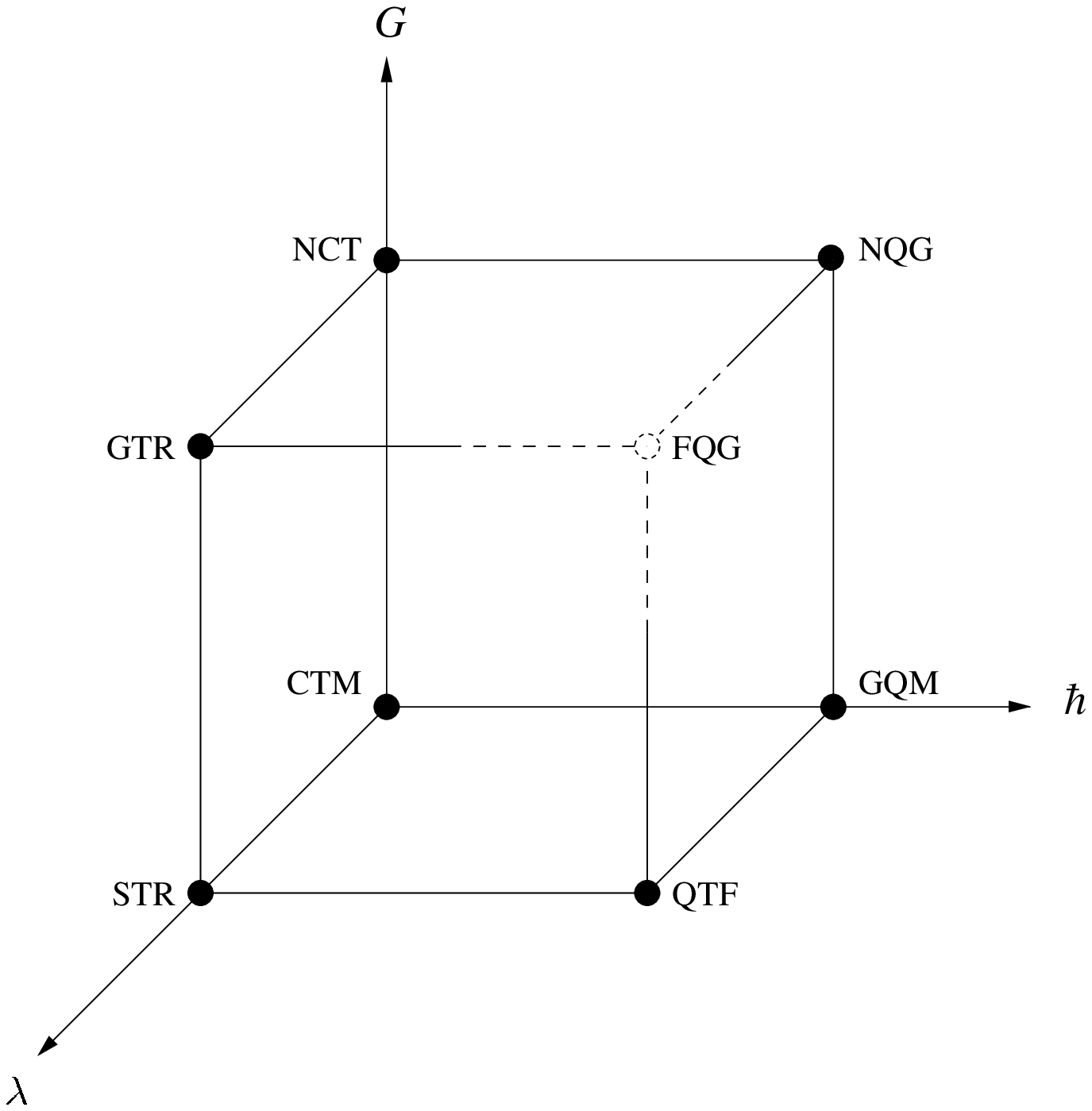}}
\vskip 0.65cm
\hrule
\bigskip
\smallskip
\parindent 0.77cm 
\baselineskip 0.5cm
{\narrower\smallskip
{\vbox to 3.35cm{\ninepoint\noindent {\bf Figure 1:}
The great dimensional monolith of physics indicating the fundamental role
played by the three universal
constants $G$ (the Newton's gravitational constant),
$\hbar$ (the Planck's constant of quanta divided by $2\pi$), and
${\lambda\equiv 1/c}$ (the `causality
constant' (Ehlers 1981),
where $c$ is the
absolute upper bound on the speed of causal influences) in various basic
theories. These theories, appearing at the eight vertices of the cube,
are: CTM = Classical Theory of Mechanics, STR = Special Theory of Relativity,
GTR = General Theory of Relativity, NCT = Newton-Cartan Theory, NQG
= Newton-Cartan Quantum Gravity (Christian 1997),
GQM = Galilean-relativistic Quantum Mechanics,
QTF = Quantum Theory of (relativistic) Fields, and FQG = the elusive Full-blown
Quantum Gravity. Note that FQG must reduce to QTF, GTR, and NQG in the
respective limits ${G\rightarrow 0}$, ${\hbar\rightarrow 0}$, and
${\lambda\rightarrow 0}$
(Kucha\v r 1980, Christian 1997, and Penrose 1997, 90-92).}}}
\endfigure

{\narrower\smallskip\noindent
``...I would like to suggest that it is possible that
quantum mechanics fails at large distances and for large objects. Now, mind
you, I do not say that I think that quantum mechanics {\it does} fail at large
distances, I only say that it is not inconsistent with what we do know. If this
failure of quantum mechanics is connected with gravity, we might speculatively
expect this to happen for masses such that ${GM^2/\hbar c=1}$, of $M$ near
${10^{-5}}$ grams, which corresponds to some ${10^{18}}$
particles.''\smallskip}

\parindent 0.00cm

Indeed, if quantum mechanics {\it does} fail near the Planck mass, as that
is the mass scale Feynman is referring to here, then -- at last -- we can
put the annoying problem of measurement to its final rest (see Figure 1 for
the meanings of the constants $G$, $\hbar$, and $c$). The judiciously
employed tool in practice, the infamous postulate often referred to as the {\it
reduction of quantum state} -- which in orthodox formulations of the theory
is taken as one of the unexplained basic postulates to resolve the tension
between the linearity of quantum dynamics and the plethora of physical
objects with apparent definite properties
-- may then be understood as an {\it objective} physical phenomenon; i.e., one
affording an ontological as opposed to epistemological understanding. From the
physical viewpoint such a resolution of the measurement problem would be
quite satisfactory, since it would render the proliferation of diverse
philosophical opinions on the matter to nothing more than a curious episode in
the history of physics. For those who are not lured by pseudo-solutions such as
the `decohering histories' approaches (Kent 1997) and/or `many worlds'
approaches (Kent 1990), a resolution of the issue by `objective reduction'
(`OR', to use Penrose's ingenious pun)
comes across as a very attractive option, provided of course that that is
indeed the path Nature has chosen to follow (see also Christian 1999a).

\parindent 0.75cm

Motivated by Feynman's inspiring words quoted above, there
have been several concrete theoretical proposals of varied sophistication
and predilections on
how the breakdown of quantum mechanics might come about such that quantum
superpositions are maintained only for `small enough' objects, whereas
reduction of the quantum state is objectively induced by gravity for
`sufficiently large' objects (K\'arolyh\'azy 1966, Komar 1969, Kibble 1981,
K\'arolyh\'azy {\it et al.} 1986, Di\'osi 1984,
1987, 1989, Ellis {\it et al.} 1989,
Ghirardi {\it et al.} 1990, Christian 1994,
Percival 1995, Jones 1995, Pearle and Squires 1996, Frenkel 1997, Fivel 1997).
Unfortunately, most of these proposals employ dubious
or {\it ad hoc} notions such as `quantum fluctuations of
spacetime' (e.g., Percival 1995) and/or
`spontaneous localization of the wavefunction' (e.g., Ghirardi {\it et al.}
1990).
Since the final `theory of everything' or `quantum gravity'
is quite far from enjoying any concrete realization (Rovelli 1998), such crude
notions cannot be relied upon when discussing
issues as fundamental as the measurement problem. In fact, these notions are
not just unreliable, but, without the context of a consistent `quantum theory
of gravity', they are also quite meaningless. For this reason, in this
essay I shall concentrate exclusively on Penrose's proposal
of quantum state reduction (1979, 1981, 1984, 1986, 1987, 1989, 367-371,
1993, 1994a, 1994b, 339-346, 1996, 1997, 1998), 
since his is a minimalist approach in which
he refrains from unnecessarily employing any ill-understood (or oxymoronic)
notions such as `quantum fluctuations of spacetime'. Rather, he argues from the
first principles, exploiting the profound and fundamental conflict (all
too familiar to anyone who has attempted to `quantize gravity' -- sometimes in
the guise of the so-called
`problem of time' (Kucha\v r 1991,
1992, Isham 1993, Belot and Earman 1999)) between
the principle of general covariance of general relativity and the principle of
superposition of quantum mechanics, to deduce a heuristic mechanism of
gravity-induced quantum state reduction. Stated differently, instead of
prematurely proposing a crude {\it theory} of quantum state reduction, he
merely provides a rationale for
the mass scale at which quantum mechanics must give way to
gravitational effects, and hence to a superior theory. 

Let me emphasize further that the motivations based on rather contentious
conceptual issues inwrought in the measurement problem are {\it not}
an essential prerequisite to Penrose's proposal for the breakdown of quantum
superpositions at a `macroscopic' scale. Instead, his proposal can
be viewed as a strategy not only to tackle the profound tension between the
foundational principles of our two most fundamental physical theories --
general relativity and quantum mechanics, but also to simultaneously provide
a possible window of opportunity to go beyond the confining principles of
these two great theories in order to arrive at even greater enveloping `final'
theory (Penrose 1984, Christian 1999b). Such a final theory, which
presumably would neither be purely quantal nor purely gravitational but
fundamentally different and superior, would then
have to reduce to quantum mechanics and general relativity, respectively,
in some appropriate approximations, as depicted in Figure 1. Clearly, unlike
the lopsided orthodox approaches towards a putative `quantum theory of
gravity' (Rovelli 1998), this is a fairly `evenhanded' approach -- as Penrose
himself often puts it. For, in the orthodox approaches, quantum superpositions
are indeed presumed to be sacrosanct at all physical scales, but only at a
very high price of some radical compromises with Einstein's theory of gravity
(e.g., at a price of having to fix both
the topological and differential structures
of spacetime {\it a priori}, as in the `loop quantum gravity'
program (Rovelli 1998), or -- even worse -- at a price of having to assume some
{\it non-dynamical} causal structure as a fixed arena for dynamical processes,
as in the currently voguish `M-theory' program (Banks 1998a, b, Polchinski
1998, Sen 1998), either of the
compromises being anathematic to the very essence of general
relativity (Einstein 1994, 155, Stachel 1994, Isham 1994, Sorkin 1997)).

        In passing, let me also point out another significant feature of
Penrose's proposal which, from a certain philosophical perspective (namely
the `process' perspective (Whitehead 1929)),
puts it in a class of very attractive proposals. Unlike some other
approaches to the philosophical problems of quantum theory, his approach (and
for that matter almost {\it all} approaches appealing to the `objective
reduction') implicitly takes {\it temporal transience} in the world -- the
incessant fading away of the specious present into the indubitable past --
not as a merely
phenomenological appearance, but as a {\it bona fide} ontological attribute of
the world, in a manner, for example, espoused by Shimony (1998). For,
clearly, any gravity-induced or other intrinsic mechanism, which purports to
actualize -- as a {\it real} physical process -- a genuine multiplicity of
quantum mechanical potentialities to a specific choice among them, evidently
captures transiency, and thereby not only goes beyond the symmetric temporality
of quantum theory, but also acknowledges the temporal transience as a
fundamental and objective attribute of the physical world (Shimony 1998)
(for anticipatory views on `becoming' along this line, see
also (Eddington 1929, Bondi 1952, Reichenbach 1956, Whitrow 1961)). A
possibility of an empirical test confirming the objectivity of this facet of
the world via Penrose's approach is by itself sufficient for me to endorse his
efforts wholeheartedly. But his approach has even more to offer. It is 
generally believed that the classical general relativistic notion of spacetime
is meaningful only at scales well above the Planck regime, and that near the
Planck scale the usual classical structure of spacetime emerges purely
phenomenologically via a phase transition or symmetry breaking
phenomenon (Isham 1994). Accordingly, one may incline to think that
``the concept of `spacetime' is not a fundamental one at all, but only
something that applies in a `phenomenological' sense when the universe is
not probed too closely'' (Isham 1997). However, if the emergence of
spacetime near the Planck scale is a byproduct of the actualization of
quantum mechanical potentialities -- via Penrose's or any related
mechanism, then the general relativistic spacetime, along with its
distinctively {\it dynamical} causal structure, comes into being not as a
coarse-grained phenomenological construct, but as a genuine ontological
attribute of the world, in close analogy with the special case of temporal
transience. In other words, such an ontological coming into being of
spacetime near the Planck scale would capture the `becoming' not merely as
temporal transience, which is a rather `Newtonian' notion, but as a much wider,
dynamical, spatio-temporal sense parallelling general relativity. (This will
become clearer in section 4 below where I discuss Penrose's mechanism, which
is tailor-made to actualize specific spacetime geometries out of
`superpositions' of such geometries).
This gratifying possibility leaves no shred of doubt that the idea of
`objective reduction' in general, and its variant proposed by Penrose in
particular, is worth investigating seriously, both theoretically and
experimentally{\parindent 0.40cm\parskip -0.50cm
\baselineskip 0.53cm\footnote{$^{\scriptscriptstyle 1}$}{\ninepoint{\hang
It should be noted that Penrose's views on `becoming' are rather different from
the stance I have taken here (1979, 1989, 1994b). In the rest of this
essay I have tried to remain as faithful to his writings as possible. For
recent discussions and further references on `becoming', other
than the paper by Shimony cited above, see
(Zeilicovici 1986), (Saunders 1996), and (Magnon 1997).\par}}}.

\parskip 0.25cm

        Since the principle of general covariance is at the heart of Penrose's
proposal, I begin in the next section with a closer look at the physical
meaning of this fundamental principle, drawing lessons from Einstein's struggle
to come to terms with it by finding a resolution of his famous `hole
argument' (1914). Even the reader fairly familiar with this episode in the
history of general relativity is urged to go through the discussion
offered here, since the subtleties of the principle of general covariance
provides the basis for both Penrose's central thesis as well as my
own partial criticism of it. Next, after highlighting the inadequacies of the
orthodox quantum measurement theory in section 3, I review Penrose's proposal
in greater detail in section 4, with a special attention to the experiment he
has proposed to corroborate his quantitative prediction of the breakdown of
quantum mechanics near a specific mass scale (subsection 4.5). (As an aside,
I also propose a new geometrical measure of gravity-induced deviation from
quantum mechanics in subsection 4.4.) Since Penrose's proposed experiment is
entirely within the nonrelativistic domain, in the subsequent subsection, 5.1,
I provide an orthodox analysis of it strictly within this domain, thereby
setting the venue for my partial criticism of his proposal in subsection 5.2.
The main conclusion here is that, since there remains no residue of
the conflict between the principles of superposition and general covariance in
the {\it strictly}-Newtonian limit (and this happens to be a rather subtle
limit), Penrose's formula for the `decay-time' of quantum superpositions
produces triviality in this limit, retaining the standard quantum coherence
intact. Finally, in subsection 5.3, before making some concluding remarks in
section 6, I suggest that an appropriate experiment which could in principle
corroborate Penrose's
predicted effect is not the one he has proposed, but a Leggett-type SQUID
(Superconducting QUantum Interference Devise) or BEC (Bose-Einstein
Condensate) experiment involving superpositions of mass distributions in
{\it relative rotation}. As a bonus, this latter analysis brings out one of
the distinctive features of Penrose's scheme, rendering it empirically
distinguishable from all of the other ({\it ad hoc}) quantum
state reduction theories involving gravity (e.g., Ghirardi {\it et al.} 1990).

\parskip 0.0cm
\parindent 0.00cm

\section{How the Spatio-temporal Events Lost Their Individuality:}

Between 1913 and 1915 Einstein (1914) put forward several versions of an
argument,
later termed by him the `hole argument' (`Lochbetrachtung'), to reject what
is known as the principle of general covariance, which he himself had elevated
earlier as a criterion for selecting the field equations of a theory of
gravitation he was in a process of constructing. It is only after two years of
struggle to arrive at the correct field equations with no avail that he was led
to reconsider general covariance, despite the hole argument, and realized the
full significance and potency of the principle it enjoys today. In particular
-- and this is also of utmost significance for our purposes here -- he realized
that the hole argument and the principle of general covariance can peacefully
coexist if and only if the mathematical individuation of the points of a
spacetime manifold is physically meaningless. In other words, he realized that
a bare spacetime manifold without some `individuating field' (Stachel 1993)
such as a specific metric tensor field defined on it is a highly fictitious
mathematical entity without any direct physical content.

\parskip 0.25cm
\parindent 0.75cm

Although the physical meaninglessness of a mathematical individuation of
spacetime points -- as a result of general covariance -- is central to
Penrose's proposal of quantum state reduction, he does not invoke the
historical episode of hole argument to motivate this nontrivial aspect of the
principle. And justifiably so. After all, the non-triviality of the
principle of general covariance (i.e., the freedom under {\it active}
diffeomorphisms of spacetime) is one of the first things one learns about while
learning general relativity. For example, Hawking and Ellis (1973) begin their
seminal treatise on the large scale structure of spacetime by
simply taking a mathematical model of spacetime to be the entire equivalence
class of copies of a 4-manifold, equipped, respectively, with Lorentzian
metric fields related by active diffeomorphisms of the manifold, without even
mentioning the hole argument. However, as we shall see, it is the hole argument
 -- an argument capable of misleading even Einstein for two years -- 
that demands such an identification in the first place. Therefore, and
especially considering the great deal of persistent confusion surrounding
the physical meaning of the principle of general covariance in the
literature (Norton 1993), for our purposes it would be worthwhile to take a
closer look at the hole argument, and thereby appreciate what is at the heart
of Penrose's proposal of quantum state reduction.
For more details on the physical meaning of general covariance the reader is
referred to Stachel's incisive analysis (1993) of it in the modern
differential geometric language; it is the general viewpoint
espoused in this reference that I shall be mostly following here (but see also
(Rovelli 1991) and
section 6 of (Anandan 1997) for somewhat analogous viewpoints).

\midfigure{hole}
\hrule
\vskip 0.95cm
\centerline{\epsfysize=7.8cm \epsfbox{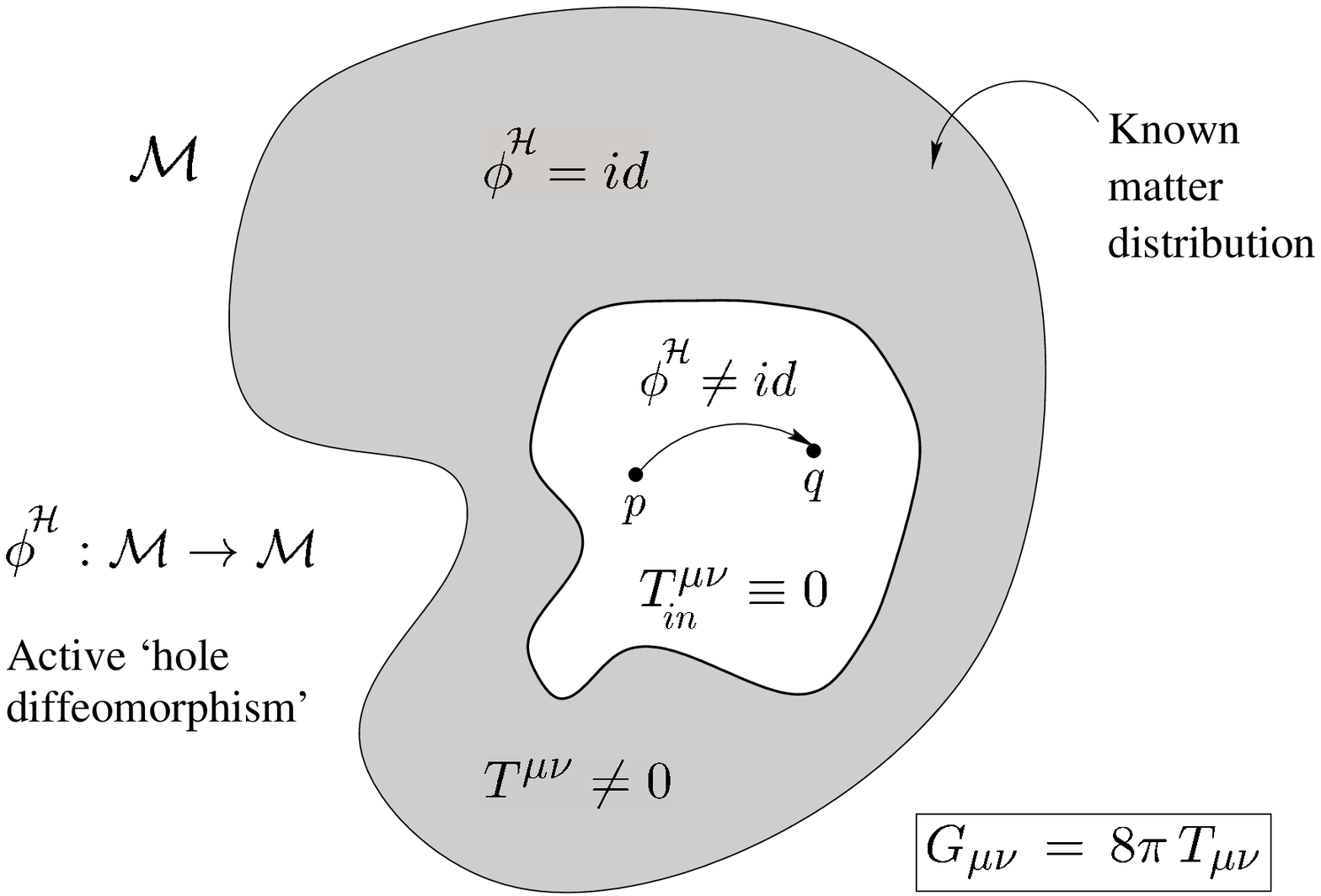}}
\vskip 0.95cm
\hrule
\bigskip
\smallskip
\parindent 0.77cm 
\baselineskip 0.5cm
{\narrower\smallskip
{\vbox to 1.0cm{\ninepoint\noindent {\bf Figure 2:}
Einstein's hole argument: If the field equations of a gravitational theory
are generally covariant, then, inside a matter-free region of some known matter
distribution, they appear to generate infinite number of inequivalent solutions
related by active diffeomorphisms of the underlying spacetime manifold.}}}
\endfigure

Without further ado, here is Einstein's hole argument:
As depicted in Figure 2, suppose that the matter distribution encoded in a
stress-energy tensor ${T^{\mu\nu}}$ is precisely known everywhere on a
spacetime ${\cal M}$ outside of some hole ${{\cal H}\subset{\cal M}}$ -- i.e.,
outside of an open subspace of the manifold ${\cal M}$. (Throughout this essay
I shall be using Penrose's abstract index notation (Wald 1984).) Further, let
there be
no physical structure defined within ${\cal H}$ except a gravitational field
represented by a Lorentzian metric tensor field ${g^{\mu\nu}}$; i.e.,
let the stress-energy vanish identically inside the hole:
${T^{\mu\nu}_{\!{}_{in}}\equiv 0}$. Now suppose that the field equations
of the gravitational theory under consideration are generally covariant.
By definition, this means that if a tensor field ${\cal X}$ on the manifold
${\cal M}$ is a solution of the set of field equations, then the pushed-forward
tensor field ${\phi_{\!{}_{*}}{\cal X}}$  of ${\cal X}$ is also a solution of
the same set of equations for any {\it active} diffeomorphism
${\phi:{\cal M}\rightarrow{\cal M}}$ of the manifold ${\cal M}$ onto itself.
The set of such diffeomorphisms of ${\cal M}$ forms a group, which is usually
denoted by ${{\rm Diff}({\cal M})}$. It is crucially important here to
distinguish between this genuine group ${{\rm Diff}({\cal M})}$ of {\it global}
diffeomorphisms of ${\cal M}$ and the pseudo-group of transformations between
overlapping pairs of local coordinate charts.
The elements of the latter group are sometimes referred to as
passive diffeomorphisms because they can only produce trivial transformations
by merely relabelling or renaming the points of a manifold. Admittance of only
tensorial objects on ${\cal M}$ in any spacetime theory is sufficient to
guarantee compatibility with this pseudo-group of passive diffeomorphisms. On
the other hand, the elements of the genuine group ${{\rm Diff}({\cal M})}$ of
active diffeomorphisms are smooth homeomorphisms
${\phi:{\cal M}\rightarrow{\cal M}}$ that can literally take each point $p$ of
${\cal M}$ into some other point ${q := \phi(p)}$ of ${\cal M}$ and thereby
deform, for example, a doughnut shaped manifold into its coffee-mug shaped
copy (Nakahara 1990, 54). Returning to the definition of general covariance,
if a metric tensor field ${g^{\mu\nu}(x)}$ is a solution of the generally
covariant field equations at any point ${x}$ of ${\cal M}$ in an adapted
local coordinate system, then so is the corresponding pushed-forward tensor
field ${(\phi_{\!{}_{*}}g)^{\mu\nu}(x)}$ at the {\it same} point $x$
in the {\it same} coordinate system. Note that, in
general, ${g^{\mu\nu}}$ and ${(\phi_{\!{}_{*}}g)^{\mu\nu}}$ will be
functionally different from each other in a given coordinate system; i.e., the
components of ${(\phi_{\!{}_{*}}g)^{\mu\nu}}$ will involve different functions
of the coordinates compared to those of ${g^{\mu\nu}}$. Now, since a choice
of ${\phi\in{\rm Diff}({\cal M})}$ 
is by definition arbitrary, nothing prevents us from choosing a smooth
${\phi^{\!{}^{\cal H}}\!}$ -- a `hole diffeomorphism' -- which reduces
to ${\phi^{\!{}^{\cal H}}\!=id}$ (i.e., identity) everywhere outside and on the
boundary of the hole ${\cal H}$, but remains ${\phi^{\!{}^{\cal H}}\!\not=id}$
within ${\cal H}$. Such a choice, owing to the fact
${T^{\mu\nu}\equiv 0}$ within the hole, implies that the action of
${\phi^{\!{}^{\cal H}}\!}$ will not affect the stress-energy tensor anywhere:
${(\phi^{\!{}^{\cal H}}_{\!{}_{*}}T)^{\mu\nu}=T^{\mu\nu}}$ everywhere, both
inside and
outside of ${\cal H}$. On the other hand, applied to the metric tensor
${g^{\mu\nu}}$, ${\phi^{\!{}^{\cal H}}}$ will of course produce a new solution
of the field equations according to the above
definition of general covariance, although outside of
${\cal H}$ this new solution will remain identical to the old solution. The
apparent difficulty, then, is that, even though ${T^{\mu\nu}}$ remains
unchanged, our choice of the hole diffeomorphism
${\phi^{\!{}^{\cal H}}\!}$ allows us to change the solution ${g^{\mu\nu}}$
inside the hole as non-trivially as we do not like, in a blatant violation of
the physically natural
uniqueness requirement, which states that the distribution of stress-energy
specified by the tensor ${T^{\mu\nu}}$ should {\it uniquely}
determine the metric tensor ${g^{\mu\nu}}$ representing the gravitational
field. Indeed, under the diffeomorphism
${\phi^{\!{}^{\cal H}}\!}$, identical matter fields ${T^{\mu\nu}}$ seam to
lead to non-trivially different gravitational fields inside the hole, such as
${g^{\mu\nu}}$ and ${(\phi^{\!{}^{\cal H}}_{\!{}_{*}}g)^{\mu\nu}}$, since
${\phi^{\!{}^{\cal H}}\!}$ is not an identity there. What is worse,
even though nothing has been allowed to change outside or on the boundary of
the hole ${\cal H}$, nothing seams to prevent the gravitational field
${(\phi^{\!{}^{\cal H}}_{\!{}_{*}}g)^{\mu\nu}}$ from being completely
different for each one of the {\it infinitely many}
inequivalent diffeomorphisms ${\phi^{\!{}^{\cal H}}\!\in{\rm Diff}({\cal M})}$
that can be carried out inside ${\cal H}$.

As mentioned above, Einstein's initial reaction to this dilemma was to abandon
general covariance for the sake of uniqueness requirement, and he
maintained this position for over two years. Of course, to a modern general
relativist a resolution of the apparent problem is quite obvious:
The tacit assumption in the hole
argument that the mathematically different tensor
fields ${g^{\mu\nu}}$ and ${(\phi^{\!{}^{\cal H}}_{\!{}_{*}}g)^{\mu\nu}}$ are
also {\it physically} different -- i.e., correspond to different physical
realities -- is clearly not justified. The two expressions
${g^{\mu\nu}}$ and ${(\phi^{\!{}^{\cal H}}_{\!{}_{*}}g)^{\mu\nu}}$ may
no matter how non-trivially differ mathematically, they must represent one and
the same gravitational field {\it physically}. Thus, as Wald puts
it (1984), ``diffeomorphisms comprise [nothing but a] gauge freedom
of any theory formulated in terms of tensor fields on a spacetime
manifold''. Accordingly, in formal analogy with the familiar gauge
freedom of the gauge field theories, modern general relativists
take a gravitational field to physically
correspond to an entire equivalence class of metric tensor fields, related by
arbitrary diffeomorphisms of the spacetime manifold,
and not just to one of the members of this class.

The analogy with the gauge freedom of the gauge field theories, however, has
only a limited appeal when it comes to general relativity. To see the
difference, recall, for example, that electromagnetic gauge transformations --
the prototype of all gauge transformations --
occur at a {\it fixed} spacetime point: The vector potential ${A_{\mu}(x)}$
defined at a point $x$ of ${\cal M}$ is physically equivalent to the vector
potential ${A_{\mu}(x)+\partial_{\mu}f(x)}$ defined at the same point $x$ of
${\cal M}$, for all scalar functions ${f(x)}$.
Although mathematically different,
both ${A_{\mu}(x)}$ and ${A_{\mu}(x)+\partial_{\mu}f(x)}$ correspond to one and
the same physical electromagnetic field configuration ${F_{\mu\nu}(x)}$, which
again depends locally on the same point $x$ of ${\cal M}$. On the other hand,
as stressed above, in general relativity diffeomorphisms
${\phi\in{\rm Diff}({\cal M})}$ map one spacetime point, say $p$, to another
spacetime point, say ${q:=\phi(p)}$. Therefore, if the tensor fields
${g^{\mu\nu}(p)}$ and ${(\phi_{\!{}_{*}}g)^{\mu\nu}(q)}$ are to be identified
as representing one and the same gravitational field configuration, implying
that they cannot be physically distinguishable by any means, then the two
points $p$ and $q$ must also be physically indistinguishable, and,
consequently,
they must renounce their individuality. For, if the points of ${\cal M}$ did
possess any ontologically significant individual identity of their own, then a
point $p$ of ${\cal M}$ could be set apart from a point $q$ of ${\cal M}$, and
that would be sufficient to distinguish the quantity ${g^{\mu\nu}(p)}$ from
the quantity ${(\phi_{\!{}_{*}}g)^{\mu\nu}(q)}$, contradicting the initial
assertion.

As Einstein eventually realized, the conclusion is inescapable: The points
of a spacetime manifold ${\cal M}$ have no direct ontological significance.
A point in a bare spacetime manifold is not distinguishable from any other
point -- and, indeed, does not even become a point (i.e, an {\it event}) with
physical meaning -- unless and until a specific metric tensor field is
{\it dynamically} determined on the manifold. In fact, in general relativity
a bare manifold not only lacks this local property, but the entire {\it global}
topological structure of spacetime is also determined only {\it a posteriori}
via a metric tensor
field (Einstein 1994, Stachel 1994, Isham 1994, Sorkin 1997). Since
a dynamical metric tensor field on a manifold dynamizes the underlying
topology of the manifold, in general relativity the topology of spacetime
is also not an absolute element that `affects without being affected'. Thus,
strictly speaking, the bare manifold does not even become `spacetime' with 
physical meaning until both the global and local spatio-temporal structures
are dynamically determined along with a metric. Further, since spacetime
points aquire
their individuality in no other way but as a byproduct of a solution of
Einstein's field equations, in general relativity `here' and `now' cannot be
part of a physical question, but can only be part of the {\it answer} to a
question, as Stachel so aptly puts it (1994). The concepts `here' and
`now' -- and hence the entire notion of local causality -- aquire ontological
meaning only {\it a posteriori}, as a part of the answer to a physical
question. Anticipating the issue discussed in the section 4 below,
this state of
affairs is in sharp contrast to what one can ask in quantum theory, which --
due to its axiomatically non-dynamical causal structure -- allows `here' and
`now' to be part of a question. Indeed, in quantum mechanics, as we shall see,
{\it a priori} individuation of spatio-temporal events is an essential
prerequisite to any meaningful notion of time-evolution.

At a risk of repetition, let me recapitulate the central point of this section
in a single sentence:

{\narrower\smallskip\noindent
{\it In Einstein's theory of gravity, general covariance --
i.e., invariance of physical laws under the action of the group}
${{\rm Diff}({\cal M})}$ {\it of} {\bf active} {\it diffeomorphisms --
expressly forbids} {\bf a priori} {\it individuation of the points of a
spacetime manifold as spatio-temporal events.}
\smallskip}

\parindent 0.00cm

Although unfairly under-appreciated (especially within approaches to
`quantum gravity' through `string' or `M' type theories, practically
{\it all} of which
being guilty of presupposing one form or another of blatantly unjustified
non-dynamical background structure
(Rovelli 1997, Banks 1998a, b, Polchinski 1998, Sen 1998, Smolin 1998)),
this is one of the most fundamental metaphysical
tenets of general relativity. In this respect, contrary to what is often
asserted following Kretschmann (1917), the principle of general
covariance is far from being physically vacuous. For instance, the potency of
general covariance is strikingly manifest in the following circumstance: if
${A_{\mu}}$ is a vector field on a general relativistic manifold ${\cal M}$,
then, unlike the situation in electromagnetism discussed above,
the value ${A_{\mu}(x)}$ at a particular point ${x\in{\cal M}}$ has no
invariant physical meaning. This is because the point $x$ can be
{\it actively} transposed around by the action of the diffeomorphism group
${{\rm Diff}({\cal M})}$, robbing it of any individuality of its own. 

\parindent 0.75cm

Of course, individuation of spacetime points
{\it can} be achieved by a {\it fixed} `gauge choice' -- that is to say, by
specification of a particular metric tensor field ${g^{\mu\nu}}$ out of the
entire equivalence class of fields ${\{\;g^{\mu\nu}\,\}}$ related by gauge
transformations, but that would be at odds with general covariance.
In fact, if there are any non-dynamical structures present, such as the
globally specified Minkowski metric tensor field ${\eta^{\mu\nu}}$ of special
relativity, then the impact of general covariance is severely
mitigated. This is because the non-dynamical
Minkowski metric tensor field, for example,
can be used to introduce a family of global inertial coordinate systems (or
`inertial individuating fields' (Stachel 1993)) that can be transformed into
each other by the (extended) Poincar\'e group of isometries of the metric:
${\hbox{\it\char36}{\!}_{\rm x}{\eta}^{\mu\nu} = 0}$, where
${\hbox{\it\char36}{\!}_{\rm x}}$ denotes the Lie derivative, with the
Killing vector
field ${{\rm x}^{\alpha}}$ being a generator of the Poincar\'e group of
transformations (Wald 1984).
These inertial coordinates in turn can be used to set apart a point $q$ from a
point $p$ of a manifold, bestowing {\it a priori} spatio-temporal individuality
to the points of the manifold (Wald 1984). For this reason,
Stachel (1993) and Wald (1984), among others, strengthen the
statement of general covariance by a condition -- explicitly added to the
usual requirement of
tensorial form for the law-like equations of physics -- that {\it there should
not be any preferred individuating fields in spacetime other than, or
independent of, the dynamically determined metric tensor field}
${g^{\mu\nu}}$. Here preferred or background fields are understood to be the
ones which affect the dynamical objects of a theory, but without being affected
by them in return. They thereby provide non-dynamical backdrops for the
dynamical processes. I shall return below to this issue of the background
structure in spacetime.

In the light of this discussion, and in response to the lack of consensus on
the meaning of general covariance in the literature (Norton 1993), let me end
this section by proposing a litmus test for general covariance --
formulated at the level of theory as a whole -- which captures its true
physical and metaphysical essence.

{\narrower\smallskip\noindent
{\bf A litmus test for general covariance:} {\it A given theory may qualify
to be called generally covariant if and only if the points of the spacetime
4-manifold, or a more general N-manifold, belonging to any model of the
theory do not possess physically meaningful {\bf a priori} individuality of
their own.}
\smallskip}

\parindent 0.00cm

(A model of a theory is a set of dynamical variables constituting a particular
solution to the dynamical equations of the theory, and may, in general, also
contain non-dynamical structures.) Admittedly, this is not a very practical
elucidation of the principle, but it does exclude theories which are not truly
generally covariant in the sense discussed above. In particular, it excludes
{\it all} of the `string' or `M' type theories known to date, since they all
presuppose individuation-condoning background structure of one form or
another. (For a recent attempt to overcome this potentially
detrimental deficiency of M-theory, see (Smolin 1998).)

\parskip 0.0cm
\parindent 0.00cm

\section{Inadequacies of the Orthodox Quantum Theory of Measurement:}

Even if we tentatively ignore the issue of individuation of spatio-temporal
events, there exist a further concern regarding the notion of definite
events in the quantum domain. In quantum theories (barring a few approaches to
`quantum gravity') one usually takes spacetime to be a fixed continuum whose
constituents are the `events' at points of space at instants of time. What is
implicit in this assumption is the classicality or definiteness of the events.
However, according to quantum mechanics, in general the notions such as `here'
and `now' could have only {\it indefinite} or {\it potential} meaning. Further,
if the conventional quantum framework is interpreted as universally applicable,
objective (i.e., non-anthropocentric), and complete (Einstein {\it et al.}
1935), then, as pointed
out above in the Introduction, the linear nature of quantum dynamics gives
rise to some serious conceptual difficulties collectively known as `the
measurement
problem' (Schr\"odinger 1935, Shimony 1963, Wheeler and Zurek 1983, Bell 1990).
These difficulties make the notion of definite or actual events in the quantum
world quite problematic, if not entirely
meaningless (Jauch 1968, Haag 1990, 1992, Shimony 1993b).
In particular, they render the
orthodox quantum theory of measurement inadequate to explain the prolific
occurrences of actual events in the `macroscopic' domain, such as the sparks
in a scintillation counter. 

\parskip 0.25cm
\parindent 0.75cm

To elucidate the measurement problem, let us consider a highly schematized 
`ideal measurement' type situation. Let ${\Sigma^{\scriptscriptstyle S}}$ and
${\Sigma^{\scriptscriptstyle A}}$ be two quantum mechanical systems 
constituting a {\it closed} composite system 
${\Sigma = \Sigma^{\scriptscriptstyle S} + \Sigma^{\scriptscriptstyle A}}$ with
their physical states represented by the rays corresponding to normalized 
vectors in the Hilbert spaces ${{\cal H}^{\scriptscriptstyle S}}$,
${{\cal H}^{\scriptscriptstyle A}}$, and 
${{\cal H}^{\scriptscriptstyle\Sigma} = 
{\cal H}^{\scriptscriptstyle S}\otimes {\cal H}^{\scriptscriptstyle A}}$, 
respectively. 
Suppose now one wants to obtain the value of a dynamical variable 
corresponding to some property of the system
${\Sigma^{\scriptscriptstyle S}}$ by means of the system
${\Sigma^{\scriptscriptstyle A}}$, which serves as a measuring apparatus. If
this dynamical variable of ${\Sigma^{\scriptscriptstyle S}}$ is represented
by a self-adjoint operator ${\Omega^{\scriptscriptstyle S}}$ in the Hilbert
space ${{\cal H}^{\scriptscriptstyle S}}$ with the eigenvalue equation
$$
\Omega^{\scriptscriptstyle S}\;\ket{\psi_j}\; = 
\;\omega_j\;\ket{\psi_j}\;\EQN eigen
$$
for some basis ${\{\ket{\psi_j}\}\in {\cal H}^{\scriptscriptstyle S}}$
(${\omega_i\not=\omega_j}$ if ${i\not=j}$), then, for 
${\Sigma^{\scriptscriptstyle A}}$ to serve the purpose of measuring the 
value of a property of ${\Sigma^{\scriptscriptstyle S}}$ in a state
${\ket{\psi_j}}$, there must be a vector 
${\ket{\varphi_{\!{}_o}}}$ in 
${{\cal H}^{\scriptscriptstyle A}}$ representing the ground state of the
system ${\Sigma^{\scriptscriptstyle A}}$ such that
after a quantum mechanical interaction between 
${\Sigma^{\scriptscriptstyle S}}$ and ${\Sigma^{\scriptscriptstyle A}}$ the
resulting final state of the composite system
${\Sigma = \Sigma^{\scriptscriptstyle S} + \Sigma^{\scriptscriptstyle A}}$ is
of the form ${\ket{\psi_j}\otimes\ket{\varphi_j}}$, where 
${\{\ket{\varphi_j}\}\in
{\cal H}^{\scriptscriptstyle A}}$ represents the set of 
indicator eigenstates
$$
Q^{\scriptscriptstyle A}\;\ket{\varphi_j}\; = 
\;q_j\;\ket{\varphi_j}\;\EQN
$$
(${q_i\not= q_j}$ if ${i\not= j}$) constituting a basis in 
${{\cal H}^{\scriptscriptstyle A}}$ with ${Q^{\scriptscriptstyle A}\in
{\cal H}^{\scriptscriptstyle A}}$ as a self-adjoint operator corresponding
to a dynamical variable representing the `indicator' property of the 
apparatus system ${\Sigma^{\scriptscriptstyle A}}$. Once such a correlation 
between the two subsystems is established, one can unequivocally infer the 
value $\omega_j$ corresponding to the property in question of the system
${\Sigma^{\scriptscriptstyle S}}$ in the state $\ket{\psi_j}$ from the 
eigenvalue $q_j$ of the indicator variable ${Q^{\scriptscriptstyle A}\,}$.
For the sake of simplicity I have assumed here that 
${\Sigma^{\scriptscriptstyle A}}$ can only be in one of a discrete, 
non-degenerate set of eigenstates ${\{\ket{\varphi_j}\}}$ 
of the indicator variable
${Q^{\scriptscriptstyle A}}$, and that the measurement interaction is of an 
`ideal' type -- i.e., the one which preserves the identity of the system
${\Sigma^{\scriptscriptstyle S}}$ of interest as well as that of the 
system ${\Sigma^{\scriptscriptstyle A}}$ treated as the measuring instrument.

So far the procedure described to infer the value of some property of a given
quantum system does not involve any ambiguity. Unfortunately, this is not the
case for more realistic initial states of the system of interest. In general,
the initial state of the system ${\Sigma^{\scriptscriptstyle S}}$ will not be 
an eigenstate but a superposition state of the form
$$
\ket{\psi}\; = \;\sum_{j=1}^N \lambda_j\;\ket{\psi_j}\;, \EQN
$$
where the scalar coefficients ${\lambda_j\in {\comps}}$, more than one being
non-zero, satisfy ${\sum_{j=1}^N |\lambda_j|^2}$ ${= 1}$ with 
${{\scriptstyle N}\equiv}$
${{\rm dim}\,{\cal H}^{\scriptscriptstyle S}}$. The post-interaction 
entangled state of the composite system
$$
\ket{\Psi}\; = \;\sum_{j=1}^N \lambda_j\;\ket{\psi_j}
\otimes\ket{\varphi_j}\EQN r
$$
dictated by the linear nature of quantum dynamics 
can now be seen to have generated an anomaly defying the very purpose of 
measurement. For, the resultant state \Ep{r} now itself is a 
superposition of ${\scriptstyle N}$ 
vectors ${\ket{\psi_j}\otimes\ket{\varphi_j}}$, each 
corresponding to a state in which the `indicator' property of the
apparatus system ${\Sigma^{\scriptscriptstyle A}}$ has a different value
$q_j\,$. In other words, the state $\ket{\Psi}$ as expressed in \Ep{r} 
implies that the `indicator' property of the apparatus system is 
{\it indefinite} (a pointer on the dial of a detector does not point
in any definite direction) obscuring the understanding of the observed
actual occurrence of events such as a formation of a droplet in a cloud
chamber, or a blackening of a silver
grain on a photographic plate. The absurdity of this direct consequence
of the linearity of quantum dynamics is well dramatized by 
Schr\"odinger in his {\it gedanken}-experiment involving
a poor cat (1935), which ends up in a limbo between
definite states of being alive and being dead.

In the conventional quantum mechanics this blatant contradiction with the 
apparent phenomenological facts about the occurrence of actual events is evaded
by invoking an {\it ad hoc} postulate --
the Projection Postulate, which in its simplest form is usually
attributed to von Neumann. According to von Neumann's theory of
measurement (1955), what has been described so far constitutes
only the first stage of measurement. The second stage involves an 
instantaneous, discontinuous and acausal change, which cannot be described 
by the usual linear and reversible quantum dynamics, and is assumed to
be accomplished by the development of 
the pure state ${W\equiv{\proj}_{\!\!{}_{\ket{\Psi}}}}$,
${\ket{\Psi}\; = \;\sum_j\lambda_j\;\ket{\psi_j}
\otimes\ket{\varphi_j}}$, into the following proper mixture
$$
W\;\longrightarrow\;W_{\!{\rm or}}\; = \;\sum_j |\lambda_j|^2\;
{\proj}_{\!\!{}_{\ket{\psi_j}\otimes\ket{\varphi_j}}}\;,\EQN PP
$$
where ${W,W_{\!{\rm or}}\in{\cal T}
({\cal H}^{\scriptscriptstyle\Sigma})^+_1}$ 
are positive normalized trace
class operators in ${{\cal H}^{\scriptscriptstyle\Sigma}}$, 
and ${{\proj}_{\!\!{}_{\ket{\eta}}}}$ 
denotes an orthonormal projector onto the one-dimensional space spanned by
a vector $\ket{\eta}$ in an appropriate Hilbert space. Note that now the
state \Ep{PP} of the composite system is consistent with, but of course
does not {\it imply}, the phenomenological Born rule:
If a quantum system ${\Sigma^{\scriptscriptstyle S}}$ in an initial pure
state represented by a unit vector ${\ket{\psi} = 
\sum_j\lambda_j\ket{\psi_j}}$ is measured by another quantum system
${\Sigma^{\scriptscriptstyle A}}$ with indicator states 
${\ket{\varphi_j}}$, then
after the measurement interaction the state of the composite system 
${\Sigma = \Sigma^{\scriptscriptstyle S}+\Sigma^{\scriptscriptstyle A}}$ is 
left in one of the pure states ${\ket{\psi_k}\otimes\ket{\varphi_k}}$ with 
probability ${|\lambda_k|^2}$, where the indicator state $\ket{\varphi_k}$ is 
correlated by the interaction with the eigenstate $\ket{\psi_k}$ of the 
dynamical variable ${\Omega^{\scriptscriptstyle S}}$ corresponding to some
property of the system ${\Sigma^{\scriptscriptstyle S}}$.

A required third and the final stage in von Neumann's scheme of measurement 
involves the contentious `ignorance interpretation of
mixtures' (Beltrametti and Cassinelli 1981, Busch {\it et al.} 1991); for 
it is not yet clear how the measuring instrument
${\Sigma^{\scriptscriptstyle A}}$ comes to exhibit a {\it definite} outcome --
i.e., how the actual occurrence of a single definite event with
corresponding relative frequency takes place
out of the compendium of various possible events encoded in the mixture
${W_{\!{\rm or}}}$. This is because the nonlinear and stochastic transition
${\ket{\Psi}\rightarrow\ket{\psi_k}\otimes\ket{\varphi_k}}$ required by the
Born rule implies the projection map ${W\rightarrow{W_{\!{\rm or}}}}$,
but {\it not} vice versa, since the mixture ${W_{\!{\rm or}}}$ in general
does not have a unique decomposition in terms of the projector
${{\proj}_{\!\!{}_{\ket{\psi_j}\otimes\ket{\varphi_j}}}}$. In general,
mixtures of numerous other projection operators may also be represented by the
same statistical operator, making it impossible to say which `basic' set of
states ${W_{\!{\rm or}}}$ is a mixture of -- a difficulty sometimes referred to
as the `preferred basis problem'. To put the rationale of difficulty in
${C^*}$-algebraic terms (Primas 1983),
the space of resultant statistical states
${W_{\!{\rm or}}}$ {\it does not} form a `simplex' in general
(Choquet and Meyer 1963),
disallowing the vectors ${\ket{\psi_j}\otimes\ket{\varphi_j}}$ to form a
`disjoint' set of vectors, and hence an ignorance or epistemic interpretation
of such mixtures is not tenable.

The upshot clearly is that von Neumann's Projection
Postulate is only a necessary but not sufficient condition for an unequivocal
understanding of the occurrence of definite events. Even if we accept this
{\it ad hoc} postulate unreservedly, the process of specific actualization out
of the compendium of quantum mechanical potentialities remains completely
obscure. Consequently,
what is desperately needed is an unequivocal {\it physical}
understanding underlying the non-unitary transition
$$
\sum_{j=1}^N \lambda_j\;\ket{\psi_j}\otimes\ket{\varphi_j}\;\;
\longrightarrow\;\;\ket{\psi_k}\otimes\ket{\varphi_k}.\EQN rrr
$$
As discussed in the Introduction above, despite a multitude of
attempts with varied sophistication and predilections, no universally
acceptable explanation -- physical or otherwise -- of this mysterious
transition is as yet in sight. In the next section we shall see that
Penrose's scheme provides precisely the much desired physical explanation
for the transition, and compellingly so.

\parskip 0.0cm
\parindent 0.00cm

\section{Penrose's Mechanism for the Objective State Reduction:}

\subsection{Motivation via a concrete example:}

To illustrate Penrose's proposal within a concrete scenario, let us apply
the above description of measurement procedure to a model interaction, within
the nonrelativistic domain, in a
specific representation -- the coordinate representation. Let us begin by
assuming a global inertial coordinate system whose origin is affixed at the
center of the earth. Using this coordinate system, let the indicator
variable ${Q^{\scriptscriptstyle A}}$ represent the location $q$ of the system
${\Sigma^{\scriptscriptstyle A}}$, which, say, has mass ${M}$, and let
the dynamical variable ${\Omega^{\scriptscriptstyle S}}$ be a time-independent
function of coordinate $x$ and its conjugate momentum 
${-i\hbar{\partial\over{\partial x}}}$ of the system
${\Sigma^{\scriptscriptstyle S}}$ exclusively.
Further, let the mass ${M}$ (i.e., the apparatus
system ${\Sigma^{\scriptscriptstyle A}}$)
be localized initially ${(t<t_{a})}$ at ${q\suba}$, and let the measurement, 
which is to be achieved by moving the mass from ${q\suba}$
to some other location, consist in the fact that if the value of
${\Omega^{\scriptscriptstyle S}(x,-i\hbar{\partial\over{\partial x}})}$ 
is ${\omega\suba}$
then the location of the mass remains unchanged at 
${q\suba}$ whereas if it is ${\omega_{j\not=
{\scriptscriptstyle 1}}}$ the mass is displaced from
${q\suba}$ to a new location ${q_{j\not={\scriptscriptstyle 1}}\equiv q\suba 
+ \omega_{j\not={\scriptscriptstyle 1}}}$.
An interaction Hamiltonian which precisely accounts for such a process 
according to the conventional Schr\"odinger equation is
(von Neumann 1955, d'Espagnat 1976)
$$
H_{int}(t) \;=\; \beta(t)\;{\widetilde{\Omega}}^{\scriptscriptstyle S}\;
P^{\scriptscriptstyle A}\;,\EQN int
$$
where $\beta(t)$ is a smooth function of time ${t\in{\real}}$
with compact support ${[t_{a},\>t_{b}]}$ satisfying 
$$
\int\limits_{t_{a}}^{t_{b}}\beta(t)\>dt\; = \;1\;,\EQN
$$ 
\offparens
$$
{\widetilde{\Omega}}^{\scriptscriptstyle S}(x,-i\hbar
{\scriptstyle{\partial\over{\partial x}}})
\;\psi_j \;=\; \{\omega_j - \omega\suba\,\delta
(\omega\suba
- \omega_j)\}\;\psi_j\;:=\; {\widetilde{\omega}}_j\;\psi_j\;,\EQN omega
$$
\autoparens
and ${P^{\scriptscriptstyle A} = -i\hbar{{\partial}\over{\partial q}}}$ with
$q$ being the indicator coordinate.
The time-dependent Schr\"odinger equation 
$$
i\hbar\,{{\partial\,}\over{\partial t}}\,
\ket{\Psi(t)}\;=\;H\!(t)\,\ket{\Psi(t)}\EQN S-eqn
$$
with the interaction Hamiltonian \Ep{int} has a general solution
$$
\Psi(t)\; = \;\psi_j\;\varphi_j(q -
q\suba - \alpha(t)\>{\widetilde{\omega}}_j)
\;,\EQN gensol
$$
where $\varphi_j$ is an arbitrary function of its argument determinable by 
initial conditions, and
$$
\alpha(t) = \int\limits_0^t \beta(t')\;dt' = 
\cases{0,&$\;\;\;\forall \;\;t < t_{a}\,$;\cr
       1,&$\;\;\;\forall \;\;t > t_{b}\,$.\cr}\EQN  
$$
For the mass ${M}$ localized at ${q\suba}$ for ${t<t_a}$, the wave-function of
the composite system ${\Sigma = \Sigma^{\scriptscriptstyle S} + 
\Sigma^{\scriptscriptstyle A}}$ is the product function
$$
\Psi(t<t_a) = \psi_j\>\delta(q - q\suba)\;,\EQN
$$
and the function $\varphi_j$ 
is identical to the delta-function. As a result, the
state after $t=t_{b}$ is
$$
\Psi(t>t_{b}) = \psi_j\>\delta(q - q\suba - {\widetilde{\omega}}_j)\EQN STA
$$
according to \Eq{gensol}. The interaction therefore induces a transition
$$
\psi_j\>\delta(q - q\suba)\;\; \mathop{\longlongrightarrow}\limits^{H_{int}}
\;\;\psi_j\>\delta(q - q\suba -{\widetilde{\omega}}_j)\;.\EQN
$$
In other words, using the definition \Ep{omega} of 
${{\widetilde{\omega}}_j}$, 
$$
\EQNalign{
\psi\suba\>\delta(q - q\suba) 
&\;\;\mathop{\longlongrightarrow}\limits^{H_{int}}\;\;
\psi\suba\>\delta(q - q\suba)\;\,
\;\;\;\;\;\;({\rm location\;of}\;M\;{\rm unchanged})\cr
{\rm but}\;\;\;\;\;
\psi_{j\not={\scriptscriptstyle 1}}\>
\delta(q - q\suba) &\;\;\mathop{\longlongrightarrow}\limits^{H_{int}}\;\; 
\psi_{j\not={\scriptscriptstyle 1}}\>
\delta(q - q_{j\not={\scriptscriptstyle 1}})\;
\;\;\;\;\;\;({\rm location\;of}\;M\;{\rm shifted}),\EQN shifted\cr}
$$
where recall that ${q_{j\not={\scriptscriptstyle 1}}
\equiv q\suba + \omega_{j\not=
{\scriptscriptstyle 1}}}$. More generally, if the 
initial state of the quantum system 
${\Sigma^{\scriptscriptstyle S}}$ is a superposition state
represented by
$$
\sum_{j=1}^N \lambda_j\,\psi_j\;,\;\;\;\;\;
\sum^N_{j=1} |\lambda_j|^2 = 1\;,\EQN mnmn
$$
then we have the above discussed
Schr\"odinger's Cat (1935) type entanglement exhibiting 
superposition of the location-states of the mass at various positions:
$$
[\sum_{j=1}^N \lambda_j\>\psi_j]\>
\delta(q - q\suba)\;\;\mathop{\longlongrightarrow}\limits^{H_{int}}\;\;
\sum_{j=1}^N \lambda_j\>\psi_j\>
\delta(q - q_j)\equiv
\sum_{j=1}^N \lambda_j\>\psi_j\>\varphi_j\;.\EQN superpos
$$
In particular, if initially we have 
$$
\Psi(t<t_a) = (\lambda\subb\psi\subb + 
\lambda\subc\psi\subc)\,\delta(q - q\suba)\,,\;\;\;\;\;
|\lambda\subb|^2+|\lambda\subc|^2=1\;,\EQN pre-limbo
$$
then, after the impulsive interaction,
$$
\Psi(t>t_b) = \lambda\subb\psi\subb\,\delta(q - q\subb)
+ \lambda\subc\psi\subc\,\delta(q - q\subc)\;,\EQN limbo
$$
and the location of the mass will be indefinite between the two 
positions ${q\subb}$ and ${q\subc}$. This of course is a perfectly respectable
quantum mechanical state for the mass ${M}$ to be in, unless it is
a `macroscopic' object and the two locations are macroscopically distinct.
In that case the indefiniteness in the location of the mass dictated by the
linearity of quantum dynamics stands in a blatant contradiction with
the evident phenomenology of such objects.

\subsection{The {\sl raison d'\^etre} of state reduction:}

Recognizing this contradiction, Penrose, among others, has tirelessly argued
that gravitation must be directly responsible for an {\it objective} resolution
of this fundamental anomaly of quantum theory (1979,
1981, 1984, 1986, 1987, 1989, 367-371,
1993, 1994a, 1994b, 339-346, 1996, 1997, 1998). 
He contends that, since the 
self-gravity of the mass must also participate in such superpositions, what
is actually involved here, in accordance with the principles of Einstein's
theory of gravity, is a superposition of two entirely {\it different}
spacetime geometries; and, when the two geometries are sufficiently different
from each other, the unitary quantum mechanical description of the
situation -- i.e., the linear superposition of a `macroscopic' mass prescribed
by \Eq{limbo} -- must breakdown{\parindent 0.40cm\parskip -0.50cm
\baselineskip 0.53cm\footnote{$^{\scriptscriptstyle 2}$}{\ninepoint{\hang
It is worth emphasising here that, as far as I can infer from his writings,
Penrose is {\it not} committed to any of the existing proposals of
nonlinear (e.g., Weinberg 1989) and/or stochastic (e.g., Pearle 1993)
modifications of quantum dynamics (neither am I for that matter).
Such proposals
have their own technical and/or interpretational problems, and are far
from being completely satisfactory. As discussed in the Introduction, Penrose's
proposal, by contrast, is truly minimalist. Rather than prematurely proposing a
{\it theory} of quantum state reduction, he simply puts forward a rationale why
his heuristic scheme for the actualization potentialities must inevitably
be a builtin feature of the sought-for
`final theory'.\par}}} (or, rather, `decay'),
allowing nature to choose between one or the other of the two geometries.

\parskip 0.25cm
\parindent 0.75cm

To understand this claim, let me surface some of the hidden assumptions
regarding spacetime structure underlying the time-evolution dictated by
\Eq{S-eqn}, which brought us to the state \Ep{limbo} in question. Recall that
I began this
section with an assumption of a globally specified inertial frame of reference
affixed at the center of earth. Actually, this is a bit too strong an
assumption. Since the Schr\"odinger equation \Ep{S-eqn} is invariant under
Galilean transformations, all one needs is a {\it family} of such global
inertial frames, each member of which is related to another by a Galilean
transformation
$$
\EQNalign{\;\;\;\;\;t\;&\longrightarrow\;\;t'\;=\;t + constant\>,\cr
          {\rm x}^{\,a}&\longrightarrow\;{\rm x'}^{\,a}\;=\;O^{a}_{\;\>b}
                               \>{\rm x}^b + \>{\rm v}^{a}\>t\> + constant\>,
                \;\;\;\;\;\;\;\;({\scriptstyle{a,b\;=\;1,2,3}}),
\EQN  Galilean\cr}
$$
where ${O^{a}_{\;\>b}\in SO(3)}$ is a time-independent orthonormal rotation
matrix (with Einstein's summation convention for like indices), and
${{\bf v}\in\real^3}$ is a time-independent spatial velocity.
Now, as discussed at the end of section 2 above, existence of a
global inertial frame grants {\it a priori} individuality
to spacetime points -- a point ${p\suba}$ of a spacetime manifold can
be set apart from a point ${p\subb}$ using such inertial
coordinates (Wald 1984, 6). Consequently, in the present scenario the concepts
`here' and `now' have {\it a priori} meaning, and they can be taken as a part
of any physical question (cf. section 2). In particular, it is meaningful to
take location ${q\suba}$ of the mass $M$ to be a part of the initial state
\Ep{pre-limbo}, since it can be set apart from any other location, such as
the location ${q\subb}$ or ${q\subc}$ in the final state \Ep{limbo}.
If individuation of spatio-temporal events was not possible, then of course
all of the locations, ${q\suba}$, ${q\subb}$, ${q\subc}$, etc., would have
been identified with each other as one and the same location, and it would not
have been meaningful to take ${q\suba}$ as a distinct initial location of the
mass (as elaborated in section 2 above, such an identification of {\it all}
spacetime points is indeed what general covariance demands in full general
relativity). Now, continuing to ignore gravity for the moment, but
anticipating Penrose's reasoning when gravity {\it is} included, let us
pretend, for the sake of argument, that the two components of the superposition
in \Eq{limbo} correspond to two {\it different} (flat) spacetime geometries.
Accordingly, let us take two separate inertial coordinate systems, one for each
spacetime but related by the transformation \Ep{Galilean}, for separately
describing the evolution of each of the two components of the superposition,
with the initial location of the mass $M$ being ${q\suba}$ as prescribed in
\Eq{pre-limbo} -- i.e., assume for the moment that each component of the
superposition is evolving on its own, as it were, under the
Schr\"odinger equation \Ep{S-eqn}. Then, for the final superposed state
\Ep{limbo} to be meaningful, a crucially important question would be: are
these two time-evolutions corresponding to the two different spacetimes
compatible with each other? In particular: is the time-translation operator
`${{\partial\,}\over{\partial\,t}}$' in \Eq{S-eqn} the same for the two
superposed evolutions -- one displacing the mass $M$ from ${q\suba}$ to
${q\subb}$ and the other displacing it from ${q\suba}$ to ${q\subc}$?
Unless the two time-translation operators in the two
coordinate systems are equivalent in some sense, we do not have a meaningful
quantum gestation of the superposition \Ep{limbo}. Now, since we are in the
Galilean-relativistic domain, the two inertial frames assigned to the two
spacetimes must be related by the transformation \Ep{Galilean}, which, upon
using the chain rule (and setting ${O\equiv\ident}$ for simplicity), yields
$$
{{\partial\;}\over{\partial\,{\rm x'}^{\,a}}}\;=\;{{\partial\;}
\over{\partial\,{\rm x}^{\,a}}}\;,\;\;\;\;\;{\rm but}\;\;\;\;\;
{{\partial\;}\over{\partial\,t'}}\;=\;{{\partial\;}
\over{\partial\,t}}\,-\,{\rm v}^a\,{{\partial\;}
\over{\partial\,{\rm x}^{\,a}}}\;.\EQN partial-v
$$
Thus, the time-translation operators are {\it not} the same for the two
spacetimes (cf. Penrose 1996, 592-593). As a result, in general, unless
${\bf v}$ identically vanishes everywhere, the two superposed time-evolutions
are {\it not} compatible with each other (see subsection 5.2, however,
for a more careful analysis). The difficulty
arises for the following reason. Although in this Galilean-relativistic domain
the individuality of spacetime points in a given spacetime is rather easy to
achieve, when it comes to two entirely {\it different} spacetimes there still
remains an ambiguity in registering the fact that the location, say
${q\subb\,}$, of the mass in one spacetime is `distinct' from its location,
say ${q\subc\,}$, in the other spacetime. On the other hand, the location
${q\subb}$ must be unequivocally distinguishable from the location ${q\subc}$
for the notion of superposition of the kind \Ep{limbo} to have any unambiguous
physical meaning. Now, in order to meaningfully set apart a location
${q\subb}$ in {\it one} spacetime from a location ${q\subc}$ in {\it another},
a point-by-point identification of the two spacetimes is clearly necessary.
But such a {\it pointwise} identification is quite ambiguous for the two
spacetimes under consideration, as can be readily seen form \Eq{Galilean},
unless the arbitrarily chosen relative spatial velocity ${\bf v}$ is set
to identically vanish everywhere (i.e., not just locally).
Of course, in the present scenario, since we have ignored gravity, nothing
prevents us from setting ${{\bf v}\equiv 0}$ everywhere -- i.e., by simply
taking nonrotating coordinate systems with constant spatial distance between
them -- and the apparent difficulty completely disappears, yielding 
$$
{{\partial\;}\over{\partial\,t'}}\;\equiv\;{{\partial\;}
\over{\partial\,t}}\;.
\EQN partial-no-v
$$
Therefore, as long as gravity is ignored, there is nothing wrong with the
quantum mechanical time-evolution leading to the superposed state \Ep{limbo}
from the initial state \Ep{pre-limbo}, since all of the hidden assumptions
exposed in this paragraph are more than justified.

\parskip 0.25cm

The situation becomes dramatically obscure, however, when one attempts
to incorporate gravity in the above scenario in full accordance with the
principles of general relativity{\parindent 0.40cm\parskip -0.40cm
\baselineskip 0.53cm\footnote{$^{\scriptscriptstyle 3}$}{\ninepoint{\hang
It is worth noting here that the conventional `quantum gravity' treatments
are of no help in the conceptual issues under consideration. Indeed, as
Penrose points out (1996, 589), the conventional attitude is to treat
superpositions of different spacetimes in merely formal fashion, in terms of
complex functions on the space of 3- or 4-geometries, with no pretence at
conceptual investigation of the physics that takes place {\it within} such a
formal superposition.\par}}}. To appreciate the central difficulty, let
us try to parallel considerations of the previous paragraph with due respect
to the ubiquitous general-relativistic features of spacetime{\parindent 0.40cm
\parskip -0.50cm
\baselineskip 0.53cm\footnote{$^{\scriptscriptstyle 4}$}{\ninepoint{\hang
Within our nonrelativistic domain, a more appropriate spacetime framework is
of course that of Newton-Cartan theory (Christian 1997). This
framework will be taken up in a later more specialized discussion, but for now,
for conceptual clarity, I rather not deviate from the subtleties of the
full general-relativistic picture of spacetime.\par}}}. To begin with,
once gravity is included, even the initial state \Ep{pre-limbo} becomes
meaningless because any location such as ${q\suba}$ loses its {\it a priori}
meaning. Recall from section 2 that in general relativity, since neither global
topological structure of spacetime nor local individuality of spatio-temporal
events has any meaning until a specific metric tensor field is dynamically
determined, the concepts `here' and `now' can only be part of the {\it answer}
to a physical question. On the other hand, the initial state \Ep{pre-limbo}
specifying the initial location ${q\suba}$ of the mass $M$ is part of the
question itself regarding the evolution of the mass. Thus, from the
general-relativistic viewpoint -- which clearly is the correct viewpoint for
a `large enough' mass -- the statement \Ep{pre-limbo} is entirely meaningless.
In practice, however, for the nonrelativistic situation under consideration,
the much more massive earth comes to rescue, since it can be used to serve as
an {\it external} frame of reference providing prior -- albeit approximate --
individuation of spacetime points (Rovelli 1991). For the sake of argument,
let us be content with such an approximate specification of the initial
location ${q\suba}$ of the mass, and ask: what role would the
general-relativistic features of spacetime play in the evolution of this
mass either from ${q\suba}$ to ${q\subb}$ or from ${q\suba}$ to ${q\subc}$,
when these two evolutions are viewed separately -- i.e., purely `classically'?
Now, since the
self-gravity of the mass must also be taken into account here, and since each
of the two evolutions would incorporate the self-gravitational effects in its
own distinct manner to determine its own overall {\it a posteriori} spacetime
geometry in accordance with the dynamical principles of Einstein's theory, to
a good degree of classical approximation there will be essentially two
{\it distinct} spacetime geometries associated with these two evolutions.
Actually, as in the case of initial location ${q\suba}$, the two final
locations, ${q\subb}$ and ${q\subc}$, would also acquire physical meaning
only {\it a posteriori} via the two resulting metric tensor fields -- say
${g\subb^{\mu\nu}}$ and ${g\subc^{\mu\nu}}$, respectively, since the
individuation of the points of each of these two spacetimes becomes meaningful
only {\it a posteriori} by means of these metric tensor fields. It is of
paramount importance here to note that, in general, the metric tensor fields
${g\subb^{\mu\nu}}$ and ${g\subc^{\mu\nu}}$ would represent two strictly
{\it separate} spacetimes with their own distinct global topological and local
causal structures. To dramatize this fact by means of a rather extreme
example, note that one of the two components of the superposition leading to
\Eq{limbo} might, in principle, end up having evolved into something like a
highly singular Kerr-Newman spacetime, whereas the other one might end up
having evolved into something like a non-singular Robertson-Walker spacetime.
This observation is crucial to Penrose's argument because, as we did in the
previous paragraph for the non-gravitational case, we must now ask whether it
is meaningful to set apart one location of the mass, say ${q\subb}$, from
another, say ${q\subc}$, in order for a superposition such as \Ep{limbo} to
have any unambiguous physical meaning. And as before, we immediately see that
in order to be able to distinguish the two locations of the mass -- i.e., to
register the fact that the mass has actually been displaced from the initial
location ${q\suba}$ to
a final location, say ${q\subb}$, and not to any other location, say
${q\subc}$ -- a point-by-point identification of the two spacetimes is
essential. However, in the present
general-relativistic picture such a pointwise identification is
{\it utterly meaningless}, especially  when the two geometries
under consideration are `significantly' different from each other. As a direct
consequence of the principle of general covariance, there is simply no
meaningful way to make a pointwise identification between two such distinct
spacetimes in general relativity. Since the theory makes no {\it a priori}
assumption as to what the spacetime manifold is and allows the Lorentzian
metric tensor field to be any solution of Einstein's field equations, the
entire causal structure associated with a general-relativistic spacetime is
dynamical and not predetermined (cf. section 2). In other words, unlike in
special relativity and the case considered in the previous paragraph, there
is simply no isometry group underlying the structure of general relativity
which could allow existence of a preferred family of inertial reference frames
that may be used, first, to individuate the points of each spacetime, and then
to identify one spacetime with another {\it point-by-point}. Furthermore, the
lack of an isometry group means that, in general, there are simply no Killing
vector fields of any kind in a general-relativistic spacetime, let alone a
time-like Killing vector field analogous to the time-translation operator
`${{\partial\,}\over{\partial\,t}}$' of the non-gravitational case considered
above (cf. \Eq{partial-v}). Therefore,
in order to continue our argument, we have to make a further assumption: We
have to assume, at least, that the spacetimes under consideration are actually
two reasonably well-defined `stationary' spacetimes with two time-like Killing
vector fields corresponding to the time-symmetries of the two metric tensor
fields ${g\subb^{\mu\nu}}$ and ${g\subc^{\mu\nu}}$, respectively. These Killing
vector fields, we hope, would generate time-translations needed to describe the
time-evolution analogous to the one provided by the
operator `${{\partial\,}\over{\partial\,t}}$' in the non-gravitational case.
However, even this drastic assumption hardly puts an end to the difficulties
involved in the notion of time-evolution leading to a superposition such as
\Ep{limbo}. One immediate difficulty is that these two Killing vector fields
generating the time-evolution are {\it completely different} for the two
components of the superposition under consideration. Since they correspond to
the time-symmetries of two essentially distinct spacetimes, they could hardly
be the same. As a result, the two Killing vector fields represent two
completely different causal structures, and hence, if we insist on implementing
them, the final state corresponding to \Eq{limbo} would involve some oxymoronic
notion such as `superposition of two distinct causalities'. Incidentally, this
problem notoriously reappears in different guises in various approaches to
`quantum gravity', and it is sometimes referred to as the `problem of
time' (Kucha\v r 1991, 1992, Isham 1993, Belot and Earman 1999). 
In summary, for a `large enough'
mass $M$, the final superposed state such as \Ep{limbo} is fundamentally and
hopelessly meaningless.

\bigskip

\parskip 0.0cm
\parindent 0.00cm

{\bf 4.3. Phenomenology of the objective state reduction:}

\parskip 0.25cm

In the previous two paragraphs we saw two extreme cases. In the first of the
two paragraphs we saw that, as long as gravity is ignored, the notion of
quantum superposition is quite unambiguous, thanks to the availability of
{\it a priori} and {\it exact} pointwise identification between the two
`spacetimes' into which a mass $M$ could evolve. However, since the ubiquitous
gravitational effects cannot be ignored for a `large enough' mass, in the last
paragraph we saw that a notion of superposition within two
general-relativistic spacetimes is completely meaningless. Thus, {\it a priori}
and {\it exact} pointwise identification of distinct spacetimes -- although
expressly forbidden by the principle of general covariance -- turns out to be
an essential prerequisite for the notion of superposition. In other words, the
superposition principle is not as fundamental a principle as the adherents of
orthodox quantum mechanics would have us believe; it makes sense only when the
other most important principle -- the principle of general covariance -- is
severely mitigated. By contrast, of course, as a result of formidable
difficulties encountered in attempts to construct a
${{\rm Diff}({\cal M})}$-invariant quantum field theory (Rovelli 1997), it
is not so unpopular to adhere that {\it active} general covariance may be truly
meaningful only at the classical general-relativistic level -- i.e., when the
superposition principle is practically neutralized.

\parindent 0.75cm

      To bridge this gulf between our two most basic principles at least
phenomenologically, Penrose invites us to contemplate an intermediate physical
situation for which the notion of quantum superposition is at best
approximately meaningful. In a nutshell, his strategy is to first
consider a `superposition' such as \Ep{limbo} with gravity included, but
nevertheless {\it a priori} pointwise identification of the two spacetimes
corresponding to the two components of the superposition retaining some
{\it approximate} meaning, and then, after putting a practical measure on this
approximation, use this measure to obtain a heuristic formula for the collapse
time of this superposition. Here is how this works: Consider two well-defined
quantum states represented by ${\ket{\Psi\subb}}$ and ${\ket{\Psi\subc}}$
(analogous to the states \Ep{STA}), each {\it stationary} on its own and
possessing the same energy $E$:
$$
i\hbar\,{{\partial\,}\over{\partial t}}\,
\ket{\Psi\subb}\;=\;E\,\ket{\Psi\subb}\;,\;\;\;\;\;\;\;
i\hbar\,{{\partial\,}\over{\partial t}}\,
\ket{\Psi\subc}\;=\;E\,\ket{\Psi\subc}\;.
\EQN Stationary
$$
In standard quantum mechanics, when gravitational effects are ignored,
linearity dictates that {\it any} superposition of these two stationary
states such as
$$
\ket{{\cal X}}\;=\;\lambda\subb\,\ket{\Psi\subb}\,+\,\lambda\subc\,
\ket{\Psi\subc}\EQN CHI
$$
(cf. \Eq{limbo}) must also be stationary, with the same energy $E$:
$$
i\hbar\,{{\partial\,}\over{\partial t}}\,
\ket{{\cal X}}\;=\;E\,\ket{{\cal X}}\;.\EQN
$$
Thus, quantum linearity necessitates a complete degeneracy of energy for
superpositions of the two original states. However, when the gravitational
fields of two different mass distributions are incorporated in the
representations ${\ket{\Psi\subb}}$ and ${\ket{\Psi\subc}}$ of these states,
a crucial question arises: will the state
${\ket{{\cal X}}}$ still remain stationary with energy $E$? Of course,
when gravity is taken into account, each of the two component states would
correspond to two entirely {\it different} spacetimes with a good degree of
classical approximation, whether or not we assume that they are reasonably
well-defined stationary spacetimes.
Consequently, as discussed above, the time-translation
operators such as `${{\partial\,}\over{\partial\,t}}$' corresponding to the
action of the time-like Killing vector fields of these two spacetimes would be
completely different form each other in general. They could only be the same
if there were an unequivocal pointwise correspondence between the two
spacetimes. Let us assume, however, that these two Killing vector fields are
not too different from each other for the physical situation under
consideration. In that case, there would be a slight -- but essential --
ill-definedness in the action of the operator
`${{\partial\,}\over{\partial\,t}}$' when it is
employed to generate a superposed state such as \Ep{CHI}, and this
ill-definedness would be without doubt reflected in the energy $E$
of this state. One can use this ill-definedness in energy, ${\Delta E}$, as a
measure of {\it instability} of the state \Ep{CHI}, and postulate the
life-time of such a `stationary' superposition -- analogous to the half-life
of an unstable particle -- to be
$$
\tau\;=\;{\hbar\over{\Delta E}}\;,\;\;\EQN The-main
$$
with two decay modes being the individual states ${\ket{\Psi\subb}}$ and
${\ket{\Psi\subc}}$ with relative probabilities ${|\lambda\subb|^2}$ and
${|\lambda\subc|^2}$, respectively. Clearly, when there is an exact pointwise
identification between the two spacetimes, ${\Delta E\rightarrow 0}$, and the
collapse of the superposition never happens. On the other hand, when such an
identification is ambiguous or impossible, inducing much larger
ill-definedness in the energy, the collapse is almost instantaneous.

A noteworthy feature of the above formula is that it is independent of the
speed of light $c\,$, implying that it remains valid even in the
nonrelativistic domain (cf. Figure 1 above and (Penrose 1994, 339, 1996, 592)).
Further, in such a Newtonian
approximation, the ill-definedness ${\Delta E}$ (for an essentially static
situation) turns out to be proportional to the gravitational self-energy of the
{\it difference} between the mass distributions belonging to the two components
of the superposition (Penrose 1996). Remarkably, numerical estimates
(Penrose 1994, 1996) based on such Newtonian models for life-times
of superpositions turn out to be strikingly realistic. For instance, the
life-time of superposition for a proton works out to be of the order of a few
million years, whereas a water droplet -- depending on its size -- is expected
to be able to maintain superposition only for a fraction of a second. Thus,
the boundary near which the reduction time is of the order of seconds is
precisely the phenomenological quantum-classical boundary of our corroborative
experience{\parindent 0.40cm\parskip -0.50cm
\baselineskip 0.53cm\footnote{$^{\scriptscriptstyle 5}$}{\ninepoint{\hang
It should be noted that, independently of Penrose, Di\'osi has also
proposed the same formula \Ep{The-main} for the collapse time (1989), but
he arrives at it from a rather
different direction. Penrose's scheme should also be
contrasted (Penrose 1996) with the `semi-classical approaches' to `quantum
gravity' (e.g., Kibble 1981), which are well-known to be
inconsistent (Eppley and Hannah 1977, Wald 1984, 382-383,
Anandan 1994). Recently, Anandan (1998) has generalized Penrose's Newtonian
expression for ${\Delta E}$ to a similar expression for an arbitrary
superposition of relativistic, but weak, gravitational fields, obtained in the
gravitational analogue of the Coulomb gauge in a linearized
approximation applied to the Lorentzian metric tensor field
(cf. subsection 5.3 for further comments).\par}}}.

\parskip 0.25cm

As I alluded to towards the end of section 3, an important issue in any quantum
measurement theory is the `preferred basis problem'. The difficulty is that,
without some further criterion, one does not know which states from the general
compendium of possibilities are to be regarded as the `basic' (or `stable' or
`stationary') states and which are to be regarded as essentially unstable
`superpositions of basic states' -- the states which are to reduce into
the basic ones. Penrose's suggestion is to regard -- within Newtonian
approximation -- the stationary solutions of what he calls the
Schr\"odinger-Newton equation as the basic states
(Penrose 1998, Moroz {\it et al.} 1998, Tod and Moroz 1998). I shall elaborate
on this equation (which I have independently
studied in (Christian 1997)) in the next section.

\bigskip

\parskip 0.0cm
\parindent 0.00cm

{\bf 4.4. A different measure of deviation from quantum mechanics:}

\parskip 0.25cm

As an aside, let me propose in this subsection
a slightly different measure for the lack of exact
pointwise identification between the two spacetimes under consideration. In
close analogy with the above assumption of stationarity, let us assume that
there exists a displacement isometry in each of the two spacetimes, embodied
in the Killing vector fields ${{\rm x}\subb}$ and ${{\rm x}\subc}$ respectively
-- i.e., let  ${\hbox{\it\char36}{\!}_{{\rm x}\subb}{g\subb}^{\mu\nu} = 0 =
{\hbox{\it\char36}{\!}_{{\rm x}\subc}{g\subc}^{\mu\nu}}}$, where
${\hbox{\it\char36}{\!}_{\rm x}}$ denotes the Lie derivative with
Killing vector fields ${{\rm x}\subb^{\alpha}}$ and ${{\rm x}\subc^{\alpha}}$
as the generators of the displacement symmetry. Further, as before, let us
assume that at least some approximate pointwise identification between these
two spacetimes is meaningful. As a visual aid, one may think of two
nearly congruent coordinate grids, one assigned to each spacetime.
Then, {\it \`a la} Penrose, I propose a measure of incongruence between these
two spacetimes to be the dimensionless parameter
${{\rm d}^{\sigma}{\rm d}_{\sigma}\,}$, taking values between zero and
unity, ${0\leq{\rm d}^{\sigma}{\rm d}_{\sigma}\leq 1}$, with
$$
{\rm d}^{\sigma}\,:=\;{\rm x}\subb^{\alpha}\nabla_{\!\alpha}\,
{\rm x}\subc^{\sigma}\;-\;{\rm x}\subc^{\alpha}\nabla_{\!\alpha}\,
{\rm x}\subb^{\sigma}\,.\EQN
$$
As it stands, this quantity is mathematically ill-defined since the Killing
vectors ${{\rm x}\subb^{\alpha}}$ and ${{\rm x}\subc^{\alpha}}$ describe the
same displacement symmetry in two quite distinct spacetimes. However, if we
reinterpret these two vectors as describing two slightly different symmetries
in one and the same spacetime, then the vector field ${{\rm d}^{\sigma}}$ is
geometrically well-defined, and it is nothing but the commutator Killing vector
field (Misner {\it et al.} 1973, 654)
corresponding to the two linearly independent vectors
${{\rm x}\subb^{\alpha}}$ and ${{\rm x}\subc^{\alpha}}$. In other words,
${{\rm d}^{\sigma}}$ then is simply a measure of incongruence
between the two coordinates adapted to simultaneously
describe symmetries corresponding to both ${{\rm x}\subb^{\alpha}}$ and
${{\rm x}\subc^{\alpha}}$ within this single spacetime. 
This measure can now be used to postulate a gravity-induced deviation
from the orthodox quantum commutation relation for the position and momentum
of the mass $M$:
$$
[Q,\,P]\;=\;i\hbar\,\{1-{\rm d}^{\sigma}{\rm d}_{\sigma}\}\,.\EQN
$$
Clearly, when there is an exact pointwise correspondence between the two
spacetimes -- i.e., when the Killing vector fields
${{\rm x}\subb^{\alpha}}$ and ${{\rm x}\subc^{\alpha}}$ are strictly
identified and ${{\rm d}^{\sigma}{\rm d}_{\sigma}\equiv 0}$,
we recover the
standard quantum mechanical commutation relation between the position and
momentum of the mass. On the other hand, when -- for a `large enough' mass --
the quantity ${{\rm d}^{\sigma}{\rm d}_{\sigma}}$ reaches order unity,
the mass exhibits essentially classical behaviour. Thus, the
parameter ${{\rm d}^{\sigma}{\rm d}_{\sigma}}$ provides a good measure of
ill-definedness in the canonical commutation relation due to a Penrose-type
incongruence, but now between the {\it displacement} symmetries of the two
spacetimes.

\bigskip

\parskip 0.0cm

{\bf 4.5. Penrose's proposed experiment:}

\parskip 0.25cm

Finally, let me end this section by describing a variant of a {\it realizable}
experiment proposed by Penrose to corroborate the contended `macroscopic'
breakdown of quantum mechanics (1998). The present version
of the experiment due to Hardy (1998) is -- arguably -- somewhat
simpler to perform. There are many practical problems in both Penrose's
original proposal and Hardy's cleverer version of it (contamination due to the
ubiquitous decoherence effects being the most intractable of all problems),
but such practical problems will not concern us here
(cf. Penrose 1998, Hardy 1998).
Further, the use of a photon in the described experiment is for convenience
only; in practice it may be replaced by any neutral particle, such as an
ultracold atom of a suitable kind.

\midfigure{Lucien}
\hrule
\vskip 0.85cm
\centerline{\epsfysize=8.3cm \epsfbox{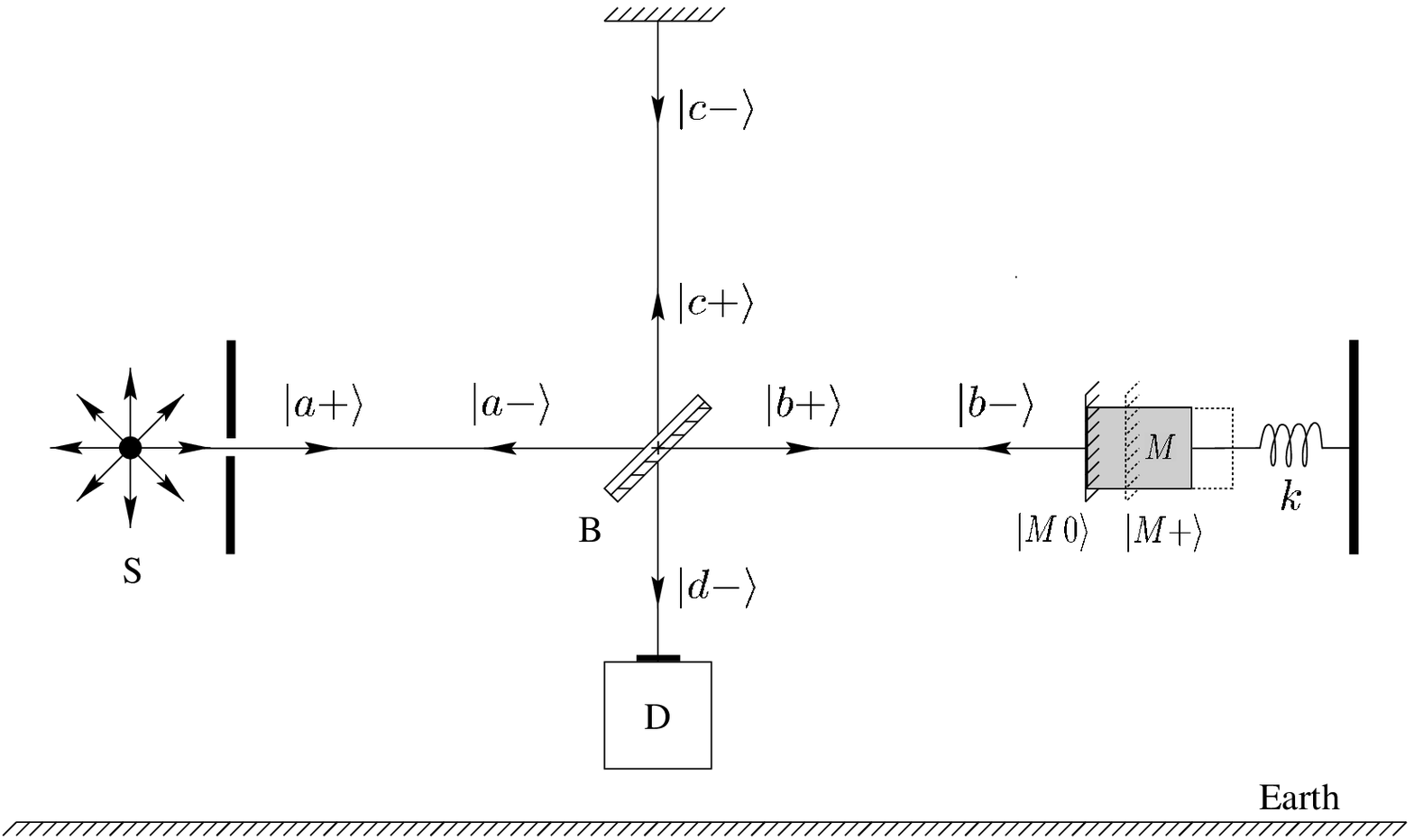}}
\vskip 0.65cm
\hrule
\bigskip
\smallskip
\parindent 0.77cm 
\baselineskip 0.5cm
{\narrower\smallskip
{\vbox to 2.2cm{\ninepoint\noindent {\bf Figure 3:}
Hardy's version of Penrose's proposed experiment: In an interferometric
arrangement, a beam-splitter, B, is placed in the `path' of an incident photon
emanating form a source S. A horizontally movable mass $M$ is attached to the
wall opposite to S by means of a restoring device with a spring
constant $k$. There are two reflecting mirrors -- one of them affixed on the
mass and the other one at the end of the vertical arm of the
interferometer, both being at an exactly equal distance from the beam splitter.
The earth provides a frame of reference, and the final destination of interest
for the photon is the detector D.}}}
\endfigure

\parindent 0.75cm

The basic experimental set-up is described in Figure 3. The system consists
of two objects: a `photon' and a `macroscopic' object of mass $M$, which in
Penrose's version is a small M\"ossbauer crystal with about ${10^{15}}$ nuclei.
The objective of the arrangement is to render the `macroscopic' mass in a
superposition of two macroscopically distinct positions, as in the state
\Ep{limbo} above. The `$+$' or `$-$' sign in the photon states (such as
${\ket{a+}}$ or ${\ket{c-}}$), respectively, indicate a forward or backward
motion along a given `path'. For now, we simply look at the arrangement in a
purely orthodox, quantum mechanical fashion. Then, the
following transformations
of the photon states due to a beam-splitter may be adopted from quantum optics:
$$
\EQNalign{\ket{a\pm}&\longleftrightarrow
\,{1\over{\sqrt 2}}\,\{\,\ket{b\pm}\,+\,\ket{c\pm}\}\cr
          \ket{d\pm}&\longleftrightarrow
\,{1\over{\sqrt 2}}\,\{\,\ket{c\pm}\,-\,\ket{b\pm}\}\,,
\EQN\cr}
$$
with inverse relations being
$$
\EQNalign{\ket{c\pm}&\longleftrightarrow
\,{1\over{\sqrt 2}}\,\{\,\ket{a\pm}\,+\,\ket{d\pm}\}\cr
          \ket{b\pm}&\longleftrightarrow
\,{1\over{\sqrt 2}}\,\{\,\ket{a\pm}\,-\,\ket{d\pm}\}\,.
\EQN\cr}
$$
If the initial state of the incident photon is taken to be ${\ket{a+}}$, and
the initial (or unmoved) state of the mass $M$ is denoted by ${\ket{M0}}$,
then the initial state of the {\it closed} composite system is the product
state
$$
\ket{a+}\otimes\ket{M0}\,.\EQN
$$
As the photon passes through the beam-splitter, this composite
initial state evolves into
$$
{1\over{\sqrt 2}}\{\,\ket{b+}\,+\,\ket{c+}\}\otimes\ket{M0}\,.\EQN
$$
Now, in the absence of the beam-splitter, if the photon happens to be in the
horizontal `path', then it would reflect off the mirror affixed on the mass,
giving it a minute momentum in the `$+$' direction. On the other hand, if the
photon is arranged to be in the vertical `path', then it would simply
reflect off the second mirror at the end of that path, without affecting
the mass. The net result of
these two alternatives in the presence of the beam-splitter, viewed quantum
mechanically, is encoded in the state
$$
{1\over{\sqrt 2}}\{\,\ket{b-}\otimes\ket{M+}\,+\,\ket{c-}\otimes\ket{M0}\}\,.
\EQN comps
$$
Since each of the two options in this superposition would lead the photon back
towards the beam-splitter, the composite state \Ep{comps} --
as the photon passes again through the beam-splitter -- will evolve into
$$
{1\over 2}\Big\lbrack\{\,\ket{a-}\,-\,\ket{d-}\}\otimes
\ket{M+}\,+\,\{\,\ket{a-}\,+\,\ket{d-}\}\otimes\ket{M0}\Big\rbrack\,.
\EQN 4.29
$$
Now, our goal here is to generate a Penrose-type superposition of the mass
$M$. Therefore, at this stage we isolate only those sub-states for
which the photon could be detected by the detector D. Thus selected from
\Ep{4.29}, we obtain
$$
{1\over{\sqrt 2}}\{\,\ket{M0}\,-\,\ket{M+}\}\EQN M-super-Penrose
$$
for the state of the mass, isolating it in the desired, spatially distinct,
`macroscopic' superposition. After some minute lapse of time, say ${\Delta t}$,
the spring will bring the mass back to its original position with its
momentum reversed, and thereby transform the above state into
$$
{1\over{\sqrt 2}}\{\,\ket{M0}\,-\,\ket{M-}\}\,,\EQN No-super-Penrose
$$
where ${\ket{M-}}$ is the new state of $M$ with its momentum in the `$-$'
direction (not shown in the figure).

At this precise moment, in order to bring about decisive statistics, we send
{\it another} photon from S into the interferometer which, upon passing
through the beam-splitter, will produce the product state
$$
{1\over 2}\,\{\,\ket{b+}\,+\,\ket{c+}\}\otimes\{\,\ket{M0}\,-\,\ket{M-}\}\,.
\EQN overa
$$
Just as before, the four terms of this state will now evolve on their
own, and, after a recoil of the photon from the two mirrors, the
composite state will become
$$
{1\over 2}\Big\lbrack\,\ket{b-}\otimes\ket{M+}\,+\,\ket{c-}\otimes\ket{M0}\,-\,
\ket{b-}\otimes\ket{M0}\,-\,\ket{c-}\otimes\ket{M-}\Big\rbrack\,.\EQN
$$
It is crucial to note here that, in the third term, the momentum of the mass
has been reduced to zero by the interaction so that both the second and
third terms have the same state ${\ket{M0}}$ for the mass. Finally, the
evolution of the photon back through the beam-splitter will render
the composite system to be in the state
$$
{1\over{2\sqrt 2}}\,\{\,\ket{a-}\,-\,\ket{d-}\}\otimes\ket{M+}\,+\,
{1\over{\sqrt 2}}\,\ket{d-}\otimes\ket{M0}\,-\,
{1\over{2\sqrt 2}}\,\{\,\ket{a-}\,+\,\ket{d-}\}\otimes\ket{M-}\,.\EQN Final
$$
Thus, quantum mechanics predicts that the probability of detecting a photon
in the detector D is ${75\%}$.

On the other hand, if the `macroscopic' superposition of the mass such as
\Ep{M-super-Penrose} has undergone a Penrose-type process of state reduction,
then the state of the mass just before the second photon is sent in
would not be \Ep{No-super-Penrose} but a {\it proper} mixture of
${\ket{M0}}$ and ${\ket{M-}}$. As a result, instead of \Ep{overa}, the overall
disjoint state after the photon has passed through the beam-splitter would
simply be
$$
{1\over{\sqrt 2}}\{\,\ket{b+}\,+\,\ket{c+}\}\otimes\ket{M0}
\;\;\;\;\;{\rm or}\;\;\;\;\;
{1\over{\sqrt 2}}\{\,\ket{b+}\,+\,\ket{c+}\}\otimes\ket{M-}\,,\EQN
$$
without any quantum coherence between the two alternatives.
As the photon is reflected off the two mirrors and passed again through the
beam-splitter, these two `classical' alternatives -- instead of \Ep{Final} --
would evolve independently into the final disjoint state
$$
{1\over{2}}\Big\lbrack\{\,\ket{a-}\,-\,\ket{d-}\}\otimes\ket{M+}\,+\,\{\,
\ket{a-}\,+\,\ket{d-}\}\otimes\ket{M0}\Big\rbrack
\;\;\;\;{\rm or}\;\;\;\;
{1\over{2}}\Big\lbrack\{\,\ket{a-}\,-\,\ket{d-}\}\otimes\ket{M0}\,+\,\{\,
\ket{a-}\,+\,\ket{d-}\}\otimes\ket{M-}\Big\rbrack\,.\EQN
$$
Consequently, if Penrose's proposal is on the right track, then, after the
photon passes through the beam-splitter second time around, it would go to the
detector only ${50\%}$ of the time and not ${75\%}$ of the time as quantum
mechanics predicts. Practical difficulties aside (Penrose 1998, Hardy 1998),
this is certainly a refutable proposition (especially because the
commonly held belief concerning decoherence (Kay 1998) -- i.e., a belief that
a strong coupling to the environment inevitably destroys the observability of
quantum effects between macroscopically distinct states -- is quite misplaced,
as emphasised by Leggett (1998)).

\parskip 0.0cm
\parindent 0.00cm

\section{A Closer Look at Penrose's Proposal within Newton-Cartan Framework:}

My main goal in this section is, first, to put forward a delicate argument
that demonstrates why Penrose's experiment -- as it stands -- is not adequate
to corroborate the signatures of his contended gravity-induced quantum
state reduction, and then to briefly discuss a couple of decisive experiments
which {\it would} be able to divulge the putative breakdown of quantum
mechanics along the line of his reasoning.

\subsection{An orthodox analysis within strictly Newtonian domain:}

In order to set the stage for my argument, let us first ask whether one can
provide an orthodox quantum mechanical analysis of the physics underlying
Penrose's proposed experiment. As it turns out, one can indeed provide
such an orthodox treatment. In this subsection I shall outline one such
treatment, which will not only direct us towards pinpointing where and
for what reasons Penrose's approach differs form the orthodox approach, but
will also allow us to explore more decisive experiments compared to
the one he has proposed.

\parskip 0.25cm
\parindent 0.75cm

Clearly, to respond to Penrose's overall conceptual scheme in orthodox mannar
one would require a full-blown and consistent quantum theory of gravity, which,
as we know, is not yet in sight (Rovelli 1998). If we concentrate, however,
not on his overall conceptual scheme but simply on his proposed experiment,
then we only require a {\it nonrelativistic} quantum theory of gravity (recall
from the last section that the formula \Ep{The-main} does not depend on the
speed of light). And, fortunately, such a theory
{\it does} exist. Recently, I have been able to demonstrate (Christian 1997)
that the covariantly described Newtonian gravity -- the so-called Newton-Cartan
gravity which duly respects Einstein's principle of equivalence -- interacting
with Galilean-relativistic matter (Schr\"odinger fields) exists as an
{\it exactly soluble} system, both classically and quantum mechanically
(cf. Figure 1). The significance of the resulting manifestly
covariant unitary quantum field theory of gravity lies in the fact that it
is the Newton-Cartan theory of gravity, and not the original Newton's theory of
gravity, that is the true Galilean-relativistic limit form of Einstein's
theory of gravity. In fact, an alternative, historically counterfactual but
logically more appropriate, formulation
of general relativity is simply Newton-Cartan theory of gravity `plus' the
light-cone structure of the special theory of relativity. Newton's original
theory in such a `generally-covariant' Newton-Cartan framework emerges
in an adscititiously chosen local inertial frame (modulo a crucially important
additional restriction on the curvature tensor, as we shall see).

To begin the analysis,
let us first look at the classical Newton-Cartan theory (for further
details and extensive references consult section II of
(Christian 1997)).
Cartan's spacetime reformulation of the classical Newtonian
theory of gravity can be motivated in exact analogy with Einstein's theory of
gravity. The analogy works because the universal equality of the inertial and
the passive gravitational masses is independent of the relativization of time,
and hence is equally valid at the Galilean-relativistic level. As a result, it
is possible to parallel Einstein's theory and reconstrue the trajectories of
(only) gravitationally affected particles as geodesics of a unique,
`{\it non-flat}' connection $\Gamma$ satisfying
$$
{{d^2x^{i}}\over{dt^2\>}} \;+\, {{\Gamma_{j\;\>k}^{\;\>i}}}\,
{{dx^{j}}\over{dt\>}}{{dx^{k}}\over{dt\>}} \;=\; 0\EQN eqa-mot
$$
in a coordinate basis, such that
$$
\Gamma_{\nu\;\>\lambda}^{\;\>\mu} \,\;\equiv \!\!\!\!\!
\buildchar{\>\;\;\;\;\Gamma_{\nu\;\>\lambda}^{\;\>\mu}}{v}{}
\,+\!\!\!\!\!\!\!
\buildchar{\>\;\;\;\;{\Theta}_{\nu\;\>\lambda}^{\;\>\mu}}{v}{} 
\;:=\!\!\!\!
\buildchar{\>\;\;\;\;\Gamma_{\nu\;\>\lambda}^{\;\>\mu}}{v}{} +\,
h^{\mu\alpha}{\!\!\buildchar{\>\;\nabla_{\!\alpha}}{v}{}}\,
{\buildchar{\Phi}{v}{}}\;t_{\nu\lambda}\;\,,\EQN gauge-depend
$$
with ${\buildchar{\Phi}{v}{}}$ representing the Newtonian gravitational
potential relative to the freely falling observer field $v$, ${\!\!\!\!\!
\buildchar{\>\;\;\;\;\Gamma_{\nu\;\>\lambda}^{\;\>\mu}}{v}{}\,}$ representing
the coefficients of the corresponding `flat' connection (i.e., one whose
coefficients can be made to vanish in a suitably chosen linear coordinate
system), and
${\!\!\!\!\!\!\!\buildchar{\>\;\;\;\;{\Theta}_{\nu\;\>\lambda}^{\;\>\mu}}{v}{}
\,:=\,h^{\mu\alpha}{\!\!\buildchar{\>\;\nabla_{\!\alpha}}{v}{}}\,
{\buildchar{\Phi}{v}{}}\;t_{\nu\lambda}\;}$ representing the traceless
gravitational field tensor associated with the Newtonian potential.
Here ${h^{\mu\nu}}$ and ${t_{\mu\nu}}$, respectively, are the degenerate
and mutually orthogonal spatial and temporal metrics with signatures
${(0\,+\,+\,+)}$ and ${(+\;0\;0\;0\,)}$, representing the
immutable chronogeometrical structure of the Newton-Cartan spacetime. They
may be viewed as the `${c\rightarrow\infty}$' limits of the Lorentzian
metric tensor field: ${h^{\mu\nu}\,=\,\lim_{c\rightarrow\infty}\,
(g^{\mu\nu}/c^2)}$ and ${t_{\mu\nu}\,=\,\lim_{c\rightarrow\infty}\,
g_{\mu\nu}}$. The conceptual superiority of this geometrization of
Newtonian gravity is reflected in the trading of the two `gauge-dependent'
quantities ${\buildchar{\Gamma}{v}{}}$ and ${\buildchar{\Theta}{v}{}}$
in favor of their gauge-independent sum ${\Gamma}$. Physically, it is the
`curved' connection ${\Gamma}$ rather than any `flat' connection
${\buildchar{\Gamma}{v}{}}$ that can be determined by local experiments. Nither
the potential ${\buildchar{\Phi}{v}{}}$ nor the `flat' connection
${\buildchar{\Gamma}{v}{}}$ has an independent existence; they exist
only relative to an arbitrary choice of a local inertial frame.
It is worth noting that, unlike in both special and general theories of
relativity, where the chronogeometrical structure of spacetime {\it uniquely}
determines its inertio-gravitational structure, in Newton-Cartan theory these
two structures are independently specified, subject only to the compatibility
conditions ${\nabla_{\!\alpha}h^{\beta\gamma}\,=\,0}$ and
${\nabla_{\!\alpha}t_{\beta\gamma}\,=\,0\,}$. In fact, the connection
${\Gamma}$, as a solution of these compatibility conditions, is not unique
unless a symmetry such as
${R^{\,\alpha\;\;\>\gamma}_{\,\;\;\>\beta\;\cdot\;\delta}\,=\,
R^{\,\gamma\;\;\>\alpha}_{\,\;\;\>\delta\;\cdot\;\beta}}$ of the curvature
tensor -- capturing the `curl-freeness' of the Newtonian gravitational filed
-- is assumed (here the indices are raised by the degenerate
spatial metric ${h^{\mu\nu}\,}$). Further, although the two metric fields are
immutable or non-dynamical in the sense that their Lie derivatives vanish
identically,
$$
\hbox{\it\char36}{\!}_{\rm x}t_{\mu\nu}\;\equiv\;0\;\;\;\;\;\;\;\;{\rm and}
\;\;\;\;\;\;\;\;
\hbox{\it\char36}{\!}_{\rm x}h^{\mu\nu}\;\equiv\;0\;,\EQN isometry
$$
the connection field remains dynamical,
${\hbox{\it\char36}{\!}_{\rm x}\,\Gamma_{\alpha\;\>\beta}^{\;\>\gamma}\,
\not=\,0}$, since it is determined by the evolving distributions of matter.
The generators ${{\rm x}=(\,t,\,{\rm x}^a)}$ of
the `isometry' group defined by the conditions
\Ep{isometry}, represented in an arbitrary reference frame, take the form
(cf. \Eq{Galilean})
$$
\EQNalign{\;\;\;\;\; t' &= \;t + constant\>,\cr
          {\rm x'}^{\,a} &= \;O^{a}_{\;\>b}(t)\>{\rm x}^b + {\rm c}^{a}\!(t)\>,
                \;\;\;\;\;\;\;\;({\scriptstyle{a,b\;=\;1,2,3}}),
\EQN Leibniz \cr}
$$
where ${O^{a}_{\;\>b}(t)\in SO(3)}$ forms an orthonormal rotation matrix for
each value of $t$ (with Einstein's summation convention for like indices),
and ${{\bf c}(t)\in\real^3}$ is an arbitrary {\it time-dependent} vector
function. Physically, these transformations connect different
observers in arbitrary (accelerating and rotating) relative motion.

With these physical motivations, the complete geometric set of gravitational
field equations of the classical Newton-Cartan theory can be written as:
$$
\EQNalign{
h^{\alpha\beta}t_{\beta\gamma}\;=\;0\,,\;\;\;\;
\nabla_{\!\alpha}h^{\beta\gamma}\;&=\;0\,,\;\;\;\;
\nabla_{\!\alpha}t_{\beta\gamma}\;=\;0\,,\;\;\;\;
\partial_{\lbrack\alpha}\,t_{\beta\rbrack\gamma}\;=\;0\;,\EQN cset; a \cr
R^{\,\alpha\;\;\>\gamma}_{\,\;\;\>\beta\;\cdot\;\delta}
&= R^{\,\gamma\;\;\>\alpha}_{\,\;\;\>\delta\;\cdot\;\beta}\;,\EQN cset; b \cr
{\rm and}\;\;\;\;\;R_{\mu\nu}\;+\;\Lambda\,t_{\mu\nu} \;&=\;
4\pi G\,M_{\mu\nu}\;,\EQN cset; c \cr} 
$$
where the first four equations specify the degenerate `metric' structure
and a set of torsion-free connections on the spacetime manifold ${\M\,}$, the
fifth one picks out the Newton-Cartan connection from this set of generic
possibilities, and the last one, with mass-momentum tensor
${M_{\mu\nu}\,:=\,\lim_{c\rightarrow\infty}\,T_{\mu\nu}}$, relates spacetime
geometry to matter in analogy with Einstein's field equations. Alternatively,
one can recover this entire set of field equations \Ep{cset} from Einstein's
theory of gravity in the `${c\rightarrow\infty}$'
limit (K\"unzle 1976, Ehlers 1981, 1986, 1991).

The only other field equation that is compatible with the structure
\Ep{cset} (Dixon 1975), but which {\it cannot} be recovered in the
`${c\rightarrow\infty}$' limit of Einstein's theory, is
$$
R^{\,\alpha\lambda\;\;\>}_{\;\;\>\;\cdot\;\gamma\delta}\;=\;0\EQN rottt
$$
(where, again, the index is raised  by the degenerate spatial metric
${h^{\lambda\sigma}\,}$). It asserts the existence of absolute rotation in
accordance with Newton's famous `bucket experiment', and turns out to be of
central importance for my argument against Penrose's experiment (cf. the next
subsection). Without this extra field equation, however, there does
not even exist a classical Lagrangian density for the Newton-Cartan system,
let alone a Hamiltonian density or an unambiguous phase space. Despite many
diligent attempts to construct a consistent Lagrangian density, the goal
remains largely elusive, thanks to the intractable geometrical
obstruction resulting from the degenerate `metric' structure of the
Newton-Cartan spacetime.

If, however, we take the condition \Ep{rottt} as an extraneously imposed but
necessary field equation on the Newton-Cartan structure, then, after some
tedious manipulations (cf. Christian 1997), we
can obtain an unequivocal constraint-free phase space for the classical
Newton-Cartan system coupled with Galilean-relativistic matter (Schr\"odinger
fields). What is more, the restriction \Ep{rottt} also permits the existence of
a family of local inertial frames in the Newton-Cartan structure (cf. the next
subsection). Given such a local frame the inertial and gravitational parts of
the Newton-Cartan connection-field can be unambiguously separated, as in the
equation \Ep{gauge-depend} above, and a non-rotating linear coordinate system
may be introduced. Then, with some gauge choices appropriate for the
earth-nucleus system of Penrose's experiment (recall that Penrose's experiment
involves displacements of some ${10^{15}}$ nuclei), the relevant action
functional (i.e., equation (4.3) of (Christian 1997)) takes the
simplified form
$$
{\cal I}\;=\,\int dt\int d\x\;[{1\over{8\pi G}}\,\Phi\,\nabla^2
\Phi\;+\;{{\,\hbar^2}\over{2m}}\,
\delta^{ab}\,\partial_a\psi\,\partial_b{\overline\psi}\;
+\;i{\hbar\over 2}(\psi\,\partial_t{\overline\psi}\,-\,{\overline\psi}\,
\partial_t\psi)\;+\;m\,{\overline\psi}\psi\,\Phi],\EQN functional
$$
where ${\psi=\psi(\x_{\scriptscriptstyle CM},\,\x)}$ is a
complex Schr\"odinger field representing the composite earth-nucleus system,
$m$ is the reduced mass for the system, all spatial derivatives are with
respect to the relative coordinate ${\x}$, and from now on the explicit
reference to observer $v$ on the top of the scalar Newtonian potential
${\Phi(\x)}$ is omitted.
Evidently, the convenient inertial frame I have chosen here is
the ${\scriptstyle CM}$-frame in which kinetic energy of the center-of-mass
vanishes identically. In addition, one may also choose
${\x_{\scriptscriptstyle CM}\equiv 0}$ without loss of generality so that
${\psi=\psi(\x)}$. Since the dynamics of the earth-nucleus system is entirely
encapsulated in the function ${\psi(\x)}$, it is sufficient to focus only
on this ${\x}$-dependence of ${\psi}$ and ignore the free motion of the
center-of-mass. Needless to say that, since ${m_{earth} \gg m_{nuleus}\,}$,
to an excellent approximation ${m = m_{nuleus}}$, and effectively the
${\scriptstyle CM}$-frame {\it is} the laboratory-frame located at the center
of the earth.

Extremization of the functional \Ep{functional} with respect to variations of
${\Phi(\x)}$ immediately yields the Newton-Poisson equation
$$
\nabla^2\Phi(\x)\;=\;-\;{4\pi G}\,m\;{\overline\psi}(\x)\,\psi(\x)
\,,\EQN Newton-N
$$
which describes the manner in which a quantum mechanically treated particle
bearing mass $m$ gives rise to a `quantized' gravitational potential
${\Phi(\x)}$, thereby capturing the essence of Newtonian quantum gravity.
On the other hand, extremization of the action with respect to variations
of the matter field ${\overline\psi(\x)}$ leads to the familiar Schr\"odinger
equation for a quantum particle of mass ${m}$ in the presence of
an external field ${\Phi(\x)}$:
$$
i\hbar{{\partial\;}\over{\partial t}}\,\psi(\x,\,t)\;=\;[-\,
{{\,\hbar^2}\over{2m}}\nabla^2\;
-\;m\,\Phi(\x)]\psi(\x,\,t)\,.\EQN Newton-M
$$
The last two equations may be reinterpreted as describing the evolution of a 
{\it single} particle of mass ${m}$ interacting with its own Newtonian
gravitational field. Then these coupled equations constitute a {\it nonlinear}
system, which can be easily seen as such by first solving equation
\Ep{Newton-N} for the potential ${\Phi(\x)}$, giving
$$
\Phi(\x)\;=\;G\,m\int d\x\,'\;
{{{\overline\psi}(\x\,')\,\psi(\x\,')}\over{|\x-\x\,'|}}\;,\EQN pot-7
$$
and then --- by substituting this solution into equation \Ep{Newton-M} ---
obtaining the integro-differential equation (cf. equation 5.18 of
(Christian 1997))
$$
i\hbar{{\partial\;}\over{\partial t}}\,\psi(\x,\,t)\,=\,
-\,{{\,\hbar^2}\over{2m}}\,\nabla^2\,\psi(\x,\,t)\,-\,G\,m^2\!\int
d\x\,'\;
{{{\overline\psi}(\x\,',\,t)\,\psi(\x\,',\,t)}\over{|\x-\x\,'|}}\;\psi(\x,\,t).
\EQN int-diff
$$
As alluded to at the end of subsection 4.3, Penrose has christened this
equation `Schr\"odinger-Newton equation', and regards the stationary solutions
of it as the `basic states' into which the quantum superpositions must reduce,
within this Newtonian approximation of the full `quantum gravity'.

As it stands, this equation is evidently a nonlinear equation describing a
self-interacting quantum particle.
However, if we promote ${\psi}$ to a `second-quantized' field operator
${\widehat\psi}$ satisfying (Christian 1997)
$$
[\,{\widehat\psi}(\x),\;{\widehat\psi}^{\dagger}\!(\x\,')]
\,=\;\widehat{\ident}\,\delta\!(\x-\x\,')\EQN ortho-1
$$
at equal-times, then this equation corresponds to a {\it linear} system of many
identical (bosonic) particles bearing mass ${m}$ in the Heisenberg picture,
with ${\widehat\psi}$ acting as an
annihilation operator in the corresponding Fock space. In particular, the
properly normal-ordered Hamiltonian operator for the system now reads
$$
\EQNalign{{\widehat{\rm H}}\;\;&\,=\,\;
{\widehat{\rm H}}\subo\;+\;{\widehat{\rm H}}_{\rm I}\;,\cr
{\rm with}\;\;\;\;\;\;\,{\widehat{\rm H}}\subo &:=\,\int d\x\;\;
{\widehat\psi}^{\dag}\!(\x)[-{{\hbar^2}\over{2m}}\nabla^2]
{\widehat\psi}(\x)\cr
{\rm and}\;\;\;\;\;\;\;{\widehat{\rm H}}_{\rm I}\,&:=\;
-\,{1\over 2}\,G\,m^2\!\int d\x\int d\x\,'\;
{{{\widehat\psi}^{\dag}\!(\x\,')\,{\widehat\psi}^{\dag}\!(\x)\,
{\widehat\psi}(\x)\,{\widehat\psi}(\x\,')}\over{|\x-\x\,'|}}\;,
\EQN Hamiltonian\cr}
$$
which upon substitution into the Heisenberg equation of motion
$$
i\hbar{{\partial\;}\over{\partial t}}\,{\widehat\psi}(\x,\,t)\;=\;
[{\widehat\psi}(\x,\,t),\,{\widehat{\rm H}}]\EQN
$$
yields an operator equation corresponding to \Ep{int-diff}. It
is easy to show (Schweber 1961, 144)
that the action of the Hamiltonian operator
${\widehat{\rm H}}$ on a multi-particle state ${\ket{\,\Psi\,}}$ is given by
$$
\bra{\x\suba\,\x\subb\,\dots\,\x_n}\,{\widehat{\rm H}}\,\ket{\,\Psi\,}=\!
[{-\,{{\,\hbar^2}\over{2m}}}\sum_{i={\scriptscriptstyle 1}}^n\nabla^2_{\!i}
\,-\,{1\over 2}\,{G\,m^2}\!\!
\sum_{\scriptstyle i,\,j=1 \atop \scriptstyle i\not=j}^n\!
{1\over{|\x_i\,-\,\x_j|}}]\!
\bra{\x\suba\,\x\subb\,\dots\,\x_n}\,\Psi\,\rangle\,,\EQN ortho-n
$$
which is indeed the correct action of the multi-particle Hamiltonian with
gravitational pair-interactions. Put differently, since the Hamiltonian
\Ep{Hamiltonian} annihilates any single-particle state, the particles no
longer gravitationally self-interact. Thus, {\it in a local inertial frame},
the Newton-Cartan-Schr\"odinger system (Christian 1997) reduces, formally, to
the very first quantum field theory constructed by Jordan and Klein (1927).

\parskip 0.0cm
\parindent 0.00cm

\subsection{The inadequacy of Penrose's proposed experiment:}

As noted above, the orthodox analysis carried out in the previous subsection
is contingent upon the extraneously imposed field equation
$$
R^{\,\alpha\lambda\;\;\>}_{\;\;\>\;\cdot\;\gamma\delta}\;=\;0\,,\EQN rottttt
$$
\Eq{rottt},
without which the existence of even a classical Lagrangian density for the
Newton-Cartan system seems impossible (cf. Christian 1997: subsection II C,
subsection IV A, footnote 6). More significantly for our purposes, unless this
extra condition prohibiting rotational holonomy is imposed on the curvature
tensor, it is not possible to recover the Newton-Poisson equation
\Ep{Newton-N},
$$
\nabla^2\Phi(\x)\;=\;-\;{4\pi G}\,\rho(\x)\,,\EQN Newton-P
$$
from the usual set of Newton-Cartan field equations \Ep{cset} (which are
obtained in the `${c\rightarrow\infty}$' limit of Einstein's theory)
without any unphysical global assumption. Thus \Ep{rottttt} embodies an
essential {\it discontinuity} in the
`${c\rightarrow\infty}$' limit between the gravitational theories of Einstein
and Newton (cf. Figure 1), and without it the Schr\"odinger-Newton equation
\Ep{Newton-M} is {\it not} meaningful.

\parskip 0.25cm
\parindent 0.75cm

Let us look at this state of affairs more closely (cf.
Misner {\it et al.} 1973,
294-295, Ehlers 1981, 1986, 1991, 1997). The only nonzero components of the
connection-field corresponding to the set of field equations \Ep{cset} (and the
coordinate transformations \Ep{Leibniz}) are
$$
\Gamma_{0\;\>0}^{\;\>a}\;=:\;-\,{\rm g}^a\;\;\;\;\;\;\;\;{\rm and}
\;\;\;\;\;\;\;\;
\Gamma_{0\;\>a}^{\;\>b}\;=\;O^{b}_{\;\>c}\,{\dot O}^{c}_{\;\>a}\;:=\;
h^{bc}\,\varepsilon_{acd}\;\Omega^d\;.\EQN Nonzero
$$
With respect to a coordinate system, the spatial vector fields
${{\bf g}(\x,\,t)}$ and ${{\bf \Omega}(\x,\,t)}$ play the part of gravitational
acceleration and Coriolis angular velocity, respectively, and the field
equations \Ep{cset} reduce to the set
$$
\EQNalign{\nabla\cdot{\bf \Omega}&=\;0\;,\;\;\;\;\;\;\;\;
\nabla\times{\bf g}\;+\;2\,{\dot{\bf \Omega}}\;=\;0\;,\cr
\nabla\times{\bf \Omega}&=\;0\;,\;\;\;\;\;\;\;\;
\nabla\cdot{\bf g}\;-\;2\,{\bf \Omega}^2\;=\;{4\pi G}\,\rho\;,\EQN\cr}
$$
where ${\bf g}$ and ${\bf\Omega}$ in general depend on both ${\x}$ and $t$
(and I have set ${\Lambda = 0}$ for simplicity).
It is clear from this set that the recovery of the Newton-Poisson equation --
and hence the reduction to the {\it strictly-Newtonian} theory -- is possible
if and only if
a coordinate system exists with respect to which ${{\bf \Omega}\,=\,0}$ holds.
This can be achieved if ${\bf\Omega}$ is spatially constant -- i.e., depends on
time only. And this is precisely what is ensured by the extra field equation
\Ep{rottttt}, which asserts that the parallel-transport of spacelike vectors
is path-independent.
Given this condition, the coordinate system can be further specialized to a
nonrotating one, with ${\,{\Gamma_{0\;\>a}^{\;\>b}}\,=\,0}$, and the connection
coefficients can be decomposed as in equation \Ep{gauge-depend},
with ${{\bf g}\,:=\,-\nabla\Phi}$.

This entire procedure, of course, may be sidestepped if we admit only
asymptotically flat spacetimes. With such a global boundary condition,
the restriction \Ep{rottttt} on the curvature tensor becomes redundant
(K\"unzle 1972, Dixon 1975). However, physical evidence
clearly suggests that we are {\it not} living in
an `island universe' (cf. Penrose 1996, 593-594) -- i.e., universe is not
`an island of matter surrounded by emptiness' (Misner {\it et al.} 1973, 295).
Therefore, a better procedure of recovering the Newtonian theory from
Einstein's theory is not to impose such a strong and unphysical global boundary
condition, but, instead, to require that only the weaker condition on the
curvature tensor, \Ep{rottttt}, is satisfied. For, this weaker condition
is quite sufficient to recover the usual version of Newton's
theory with gravitation as a force field on a flat, non-dynamical,
{\it a priori} spacetime structure, and
guarantees existence of a class of inertial coordinate systems {\it not}
rotating with respect to each other; i.e., the condition suppresses
time-dependence of the rotation matrix ${O^{a}_{\;\>b}(t)}$ (as
a result of the restriction ${\,{\Gamma_{0\;\>a}^{\;\>b}}\,=\,0}$), and
reduces the transformation law \Ep{Leibniz} to 
$$
\EQNalign{\;\;\;\;\;t\;&\longrightarrow\;\;t'\;=\;t + constant\>,\cr
           {\rm x}^{\,a}&\longrightarrow\;{\rm x'}^{\,a}\;=\;
                O^{a}_{\;\>b}\>{\rm x}^b + {\rm c}^{a}\!(t)\>,
                \;\;\;\;\;\;\;\;({\scriptstyle{a,b\;=\;1,2,3}})\,.
\EQN Leibniz-no-time \cr}
$$
Note that, unlike the asymptotic-flatness imposing condition
${\lim_{{|\x|}\to\infty}\Phi(\x)\,=\,0}$, the weaker condition \Ep{rottttt}
{\it does not} suppress the arbitrary time-dependence of the function
${{\rm c}^{a}(t)}$ -- i.e., \Ep{rottttt} does not reduce ${{\rm c}^{a}(t)}$ to
${{\rm v}^{a}\times t}$ as in the Galilean transformation \Ep{Galilean} above.
Consequently, the gravitational potential ${\Phi}$ in the resultant Newtonian
theory remains nonunique (Misner {\it et al.} 1973, 295), and, under the
diffeomorphism corresponding
to the transformation \Ep{Leibniz-no-time}, transforms (actively) as
$$
\Phi(\x)\;\longrightarrow\;\Phi'(\x)\;=\;\Phi(\x)\;-\;
{\ddot{\bf c}}\cdot\x\,.\EQN Phi-transformation
$$

Let us now go back to Penrose's hypothesis on the mechanism underlying
quantum state reduction discussed in the subsection 4.2 above, and
retrace the steps of that subsection within the present strictly-Newtonian
scenario. As before, although here ${h^{\mu\nu}}$ and ${t_{\mu\nu}}$ would
serve as `individuating fields' (cf. section 2) allowing pointwise
identification between two different spacetimes, due to the transformation law
\Ep{Leibniz-no-time} there would appear to be an ambiguity in the notion of
time-translation operator analogous to \Eq{partial-v},
$$
{{\partial\;}\over{\partial\,{\rm x'}^{\,a}}}\;=\;{{\partial\;}
\over{\partial\,{\rm x}^{\,a}}}\;\;\;\;\;\;{\rm but}\;\;\;\;\;
{{\partial\;}\over{\partial\,t}}\;\longrightarrow\;
{{\partial\;}\over{\partial\,t'}}\;=\;{{\partial\;}
\over{\partial\,t}}\,-\,{\rm\dot c}^a(t)\,{{\partial\;}
\over{\partial\,{\rm x}^{\,a}}}\;,\EQN partial-time-v
$$
when superpositions involving two such different spacetimes are considered.
However, I submit that this `ambiguity' in the present -- essentially Newtonian
-- case is entirely innocuous. For, in the strictly Newtonian theory being
discussed here, where a `spacetime' now is simply a flat structure `plus' a
gravitational potential ${\Phi(\x)}$ as in equation \Ep{gauge-depend}, one must
consider \Ep{partial-time-v} {\it together} with the transformation
\Ep{Phi-transformation}. But the Schr\"odinger-Newton equation \Ep{Newton-M} --
which is the appropriate equation here -- happens to be covariant under such a
concurrent transformation{\parindent 0.40cm\parskip -0.50cm
\baselineskip 0.53cm\footnote{$^{\scriptscriptstyle 6}$}{\ninepoint{\hang
Better still: under simultaneous gauge transformations
\Ep{Phi-transformation}, \Ep{partial-time-v} and \Ep{brown-phase}, the
Lagrangian density of the action \Ep{functional} remains
invariant except for a change in the spatial boundary term, which of course
does not contribute to the Euler-Lagrange equations \Ep{Newton-N} and
\Ep{Newton-M}.
Thus, the entire Schr\"odinger-Newton theory is unaffected by these
transformations, implying that it is independent of a particular
choice of reference frame represented by ${{\partial\;\over{\partial t}}}$ out
of the whole family given in \Ep{partial-time-v}. It should be noted, however,
that here, as in any such demonstration of covariance, all variations
${\delta\Phi}$ of the Newtonian potential is assumed to
vanish identically at the spatial boundary.\par}}}, and
retains the original form
$$
i\hbar{{\partial\;}\over{\partial t'}}\,\psi'(\x,\,t)\;=\;[-\,
{{\,\hbar^2}\over{2m}}\nabla'^2\;
-\;m\,\Phi'(\x)]\psi'(\x,\,t)\EQN Newton-M
$$
(Rosen 1972, cf. also Christian 1997)
with the following (active) transformation of its solution (if it exists):
$$
\psi(\x,\,t)\;\longrightarrow\;\psi'(\x,\,t)\;=\;e^{if(\x,\,t)}\,\psi(\x,\,t)
\,.\EQN brown-phase
$$
What is more (cf. Kucha\v r 1980, 1991), due to inverse relation between
transformations on the function space and transformations \Ep{Leibniz-no-time}
on coordinates, equation \Ep{brown-phase} implies
$$
\psi'(\x',\,t')\;=\;\psi(\x,\,t)\;.\EQN psi-equivalent
$$
That is to say, the new solution of the Schr\"odinger-Newton equation expressed
in the new coordinate system is {\it exactly}
equal to the old solution expressed in
the old coordinate system -- the new value of the ${\psi}$-field, as measured
at the transformed spacetime point, is numerically the same as its old value
measured at the original spacetime point. Now consider a superposition
involving two entirely different strictly-Newtonian `spacetimes' in the
coordinate representation analogous to the `superposition' \Ep{CHI} discussed
in section 4,
$$
\langle\x\ket{{\cal X}(t)}\;=\;\lambda\subb\,
\Psi\subb(\x,\,t)\,+\,\lambda\subc\,\Psi'\subc(\x',\,t')\;,\EQN new-CHI
$$
where unprimed coordinates correspond to one spacetime and the primed
coordinates to another{\parindent 0.40cm\parskip -0.50cm
\baselineskip 0.53cm\footnote{$^{\scriptscriptstyle 7}$}{\ninepoint{\hang
Of course, since the Schr\"odinger-Newton equation is a non-linear equation,
its more adequate (orthodox) quantum-mechanical treatment is the one given
by equations \Ep{ortho-1}--\Ep{ortho-n} of the previous subsection. My purpose
here, however, is simply
to parallel Penrose's argument of instability in quantum superpositions
near the Planck mass.\par}}}.
{\it Prima facie}, in accordance with the reasonings
of section 4, such a superposition should be as unstable as \Eq{CHI}. However,
in the present strictly-Newtonian case, thanks to the relation
\Ep{psi-equivalent}, the physical state represented by \Ep{new-CHI} is
equivalent to the superposed state
$$
\langle\x\ket{{\cal X}(t)}\;=\;\lambda\subb\,
\Psi\subb(\x,\,t)\,+\,\lambda\subc\,\Psi\subc(\x,\,t)\;.\EQN old-CHI
$$
And there is, of course, nothing unstable about such a superposition in this
strictly-Newtonian domain. Consequently, for such a superposition,
${{\Delta E}\equiv 0}$, and hence its life-time
${\tau\sim\infty}$ (cf. \Eq{The-main}).

Thus, as long as restriction \Ep{rottttt} on the curvature tensor is
satisfied -- i.e., as long as it is possible to choose a coordinate system
with respect to which ${\,{\Gamma_{0\;\>a}^{\;\>b}}\,=\,0}$ holds for each
spacetime, the
Penrose-type instability in quantum superpositions is non-existent (a
conclusion not inconsistent with the results of (Christian 1997)). Put
differently, given ${\,{\Gamma_{0\;\>a}^{\;\>b}}\,=\,0}$, the Penrose-type
obstruction to stability of superpositions is sufficiently mitigated to sustain
stable quantum superpositions. In physical
terms, since \Ep{rottttt} postulates the existence of `absolute rotation',
the superposition \Ep{new-CHI} is perfectly Penrose-stable as long as there
is no {\it relative} rotation involved between its two components. On the other
hand, if there {\it is} a relative rotation between the two components of
\Ep{new-CHI} so that ${\,{\Gamma_{0\;\>a}^{\;\>b}}\,=\,0}$ does not hold for
both spacetimes, than it is not possible to analyze the physical system in
terms of the strictly-Newtonian limit of Einstein's theory, and, as a result,
the `superposition' \Ep{new-CHI} would be Penrose-unstable.
Unfortunately, neither in Penrose's original experiment (1998), nor in the
version discussed in subsection 4.5 above, is there any relative rotation
between two components of the superposed mass distributions. In other
words, in both cases ${\,{\Gamma_{0\;\>a}^{\;\>b}}\,=\,0}$ holds everywhere,
and hence no Penrose-type instability should be expected in the outcome of
these experiments. (Incidentally, among the known solutions of Einstein's field
equations, the only known solution which has a genuinely Newton-Cartan limit
-- i.e., in which ${\bf\Omega}$ is not spatially constant, entailing that it
cannot be reduced to the strictly-Newtonian
case with ${\,{\Gamma_{0\;\>a}^{\;\>b}}\,=\,0\,}$ -- is the NUT spacetime
(Ehlers 1997)).

\parskip 0.0cm
\parindent 0.00cm

\subsection{More adequate experiments involving relative rotations:}

It is clear from the discussion above that, in order to detect Penrose-type
instability in superpositions, what we must look for is a physical system
for which the components ${\Gamma_{0\;\>a}^{\;\>b}}$ of the connection field,
{\it in addition} to the components ${\Gamma_{0\;\>0}^{\;\>a}\,}$, are
meaningfully non-zero.
Most conveniently, there exists extensive theoretical and experimental work
on just the kind of physical systems we require. 

\parskip 0.25cm
\parindent 0.75cm

The first among these systems involves
`macroscopic' superpositions of two screening currents in r.f.-SQUID rings,
first proposed by Leggett almost two decades ago (Leggett 1980, 1984, 1998,
Leggett and Garg 1985). An r.f.-SQUID ring consists of a loop of
superconducting
material interrupted by a thin Josephson tunnel junction. A persistent
screening current may be generated around the loop in response to an externally
applied magnetic flux, which obeys an equation of motion similar to that of a
particle moving in a one dimensional double-well potential. The thus generated
current in the ring would be equal in magnitude in both wells, but opposite in
direction. If dissipation in the junction and decoherence due to environment
are negligible, then the orthodox quantum analysis predicts coherent
oscillations between the two distinct flux states, and, as a result, a coherent
superposition between a large number of electrons flowing around the ring in
opposite directions -- clockwise or counterclockwise -- is expected to exist,
generating a physical situation analogous to the one in \Eq{CHI} or
\Ep{new-CHI} above. Most importantly for our purposes, since there would be
relative rotation involved between the currents in the two possible states,
owing to the Lense-Thirring fields (Lense and Thirring 1918,
Ciufolini {\it et al.} 1998) of these currents, the connection components
${\Gamma_{0\;\>a}^{\;\>b}}$, in addition to the components
${\Gamma_{0\;\>0}^{\;\>a}}$, will be nonzero. And this will unambiguously give
rise to a Penrose-type instability at an appropriate mass scale -- say roughly
around ${10^{21}}$ electrons. The number of electrons in the SQUID ring in an
actual experiment currently under scrutiny in Italy (Castellano {\it et al.}
1996) is only of the order of ${10^{15}}$, but there is no reason for a
{\it theoretical} upperbound on this number.

It should be noted that Penrose himself has briefly considered the possibility
of a Leggett-type experiment to test his proposal (1994b, 343). Recently,
Anandan (1998) has generalized Penrose's expression for ${\Delta E}$ to
arbitrary connection fields (cf. footnote 5), which allows him to
consider connection components other than ${\Gamma_{0\;\>0}^{\;\>a}}$,
in particular the components ${\Gamma_{0\;\>a}^{\;\>b}}$,
and suggest a quantitative test of Penrose's ansatz via Leggett's experiment.
What is novel in my own endorsement of this suggestion is the realization that
Leggett-type experiments belong to a class of experiments -- namely, the class
involving ${\Gamma_{0\;\>a}^{\;\>b}\not=0}$ -- which is the only class
available within the nonrelativistic domain to unequivocally test Penrose's
proposal.

A second more exotic physical system belonging to this class of experiments
is a superposition of two vortex states of
an ultracold Bose-Einstein Condensate (BEC), currently being studied by Cirac's
group in Austria among others (Cirac {\it et al.} 1998, Dum {\it et al.} 1998,
Butts and Rokhsar 1999).
Again, owing to the Lense-Thirring fields of such a slowly whirling BEC
(clockwise or counterclockwise), a Penrose-type instability can in principle be
detected at an appropriate mass scale.

Finally, let me point out that the analysis of this section has opened up an
exciting new possibility of empirically distinguishing Penrose's scheme from
other ({\it ad hoc}) theories of gravity-induced state reduction
(e.g., Ghirardi {\it et al.} 1990), with the locus of differentiation being
the connection components ${\Gamma_{0\;\>a}^{\;\>b}}$. There is nothing
intrinsic in such {\it ad hoc} theories that could stop a state from reducing
when these connection components are zero -- e.g., for the experiment described
in subsection 4.5 above these theories predict reduction at an appropriate
scale, whereas Penrose's scheme, for the reasons explicated above, {\it does
not}.

\parskip 0.0cm
\parindent 0.00cm

\section{Concluding Remarks:}

Notwithstanding the importance of partial reservations levelled against
Penrose's proposed experiment in the previous section, it should be clear that
my criticism has significance only in the strictly-Newtonian domain.
The classical world, of course, is not governed by Galilean-relativistic
geometries, but by general-relativistic geometries. Accordingly, the true
domain of the discussion under consideration must be the domain of full
`quantum gravity'. And, reflecting on this domain, I completely share
Penrose's sentiments that ``our present picture of physical reality,
particularly in relation to the nature of {\it time}, is due for a grand
shake up'' (1989, 371) (similar sentiments, arrived at
from quite a different direction, are also expressed by Shimony (1998)).
The incompatibility between the fundamental principles of our two most basic
theories -- general relativity and quantum mechanics -- is so severe that the
unflinching orthodox view maintaining a {\it status quo} for quantum
superpositions -- including at such a special scale as the Planck scale -- is
truly baffling. As brought out in several of the essays in this collection and
elaborated by myself in section 4 above, the
conflict between the two foundational theories has primarily to do with the
axiomatically presupposed fixed causal structure underlying quantum dynamics,
and the meaninglessness of such a fixed, non-dynamical, background causal
structure in the general relativistic picture of the world. The orthodox
response to the conflict is to hold the fundamental principles of quantum
mechanics absolutely sacrosanct at the price of severe compromises with those
of Einstein's theory of gravity. For example, Banks, one of the pioneers of
the currently popular M-theory program, has proclaimed (1998b): `` ... it seems
quite clear that the fundamental rules of [M-theory] will seem outlandish to
anyone with a background in ... general relativity. ... At the moment it
appears that the only things which may remain unscathed are the fundamental
principles of quantum mechanics.'' In contrast, representing a view of growing
minority, Penrose has argued for a physically more meaningful {\it evenhanded}
approach in which even the superposition principle is not held beyond
reproach at all scales. It certainly requires an extraordinary leap of faith
in quantum mechanics (a leap, to be precise, of some {\it seventeen} orders of
magnitude in the length scale!) to maintain that the Gordian knot -- the
conflict between our two most basic theories -- can be cut without compromising
the superposition principle in some manner. My own feeling, heightened by
Penrose's tenacious line of reasoning,
is that such a faith in quantum mechanics could turn out to
be fundamentally misplaced, as so tellingly made plain by Leggett (1998):

\smallskip

\parskip 0.25cm
\parindent 0.55cm

{\narrower\smallskip\noindent
``Imagine going back to the year 1895 and telling one's colleagues that
classical mechanics would break down when the product of energy and time
reached a value of order ${10^{-34}}$ joule seconds. They would no doubt
respond gently but firmly that any such idea must be complete nonsense, since
it is totally obvious that the structure of classical mechanics cannot tolerate
any such characteristic scale!''\smallskip}

\smallskip

\parindent 0.00cm

Indeed, one often comes across similar sentiments with regard to the beautiful
internal coherence of quantum formalism. However, considering the extraordinary
specialness of the Planck scale, I sincerely hope that our `quantum' colleagues
are far less complacent than their `classical' counterparts while harbouring
the `dreams of a final theory'.

\smallskip
\parskip 0.0cm
\parindent 0.0cm

\nosechead{\tbf Acknowledgements}

I am truly grateful to my mentor Abner Shimony for his kind and generous
financial support without which this work would not have been possible.
I am also grateful to Roger Penrose for discussions on his ideas about
gravity-induced state reduction, Lucien Hardy for kindly letting me
use his own version of Penrose's proposed experiment before publication,
Ashwin Srinivasan for his expert help in casting figures in TeX, and Jeeva
Anandan, Julian Barbour, Harvey Brown, Roger Penrose and Paul Tod for their
comments on parts of the manuscript.

\vfil\eject

\parskip 0.12cm
\parindent -0.85cm
\hoffset=0.50cm

{\tbf References}

\bigskip

Anandan, J. S. (1994), ``Interference of Geometries in Quantum Gravity'',
{\underbar{General Relativity and Gravitation}} {\underbar{26}}: 125-133.

Anandan, J. S. (1997), ``Classical and Quantum Physical Geometry'', in
R. S. Cohen, M. Horn, and J. Stachel (eds.),
{\underbar{Potentiality, Entanglement and Passion-at-a-Distance:
Quantum Mechanical Studies for Abner}}\n {\underbar{Shimony, Volume Two}}. 
Dordrecht: Kluwer Academic, pp. 31-52.

Anandan, J. S. (1998), ``Quantum Measurement Problem and the Gravitational
Field'', in S. A. Huggett, L. J. Mason, K. P. Tod, S. T. Tsou, and N. M. J.
Woodhouse (eds.),
{\underbar{The Geometric Universe: Science, Geometry, and}} {\underbar{the Work
of Roger Penrose}}. Oxford, England: Oxford University Press, pp. 357-368
[see also gr-qc/9808033].

Banks, T. (1998a), ``Matrix Theory'', {\underbar{Nuclear Physics
B -- Proceedings Supplements}} {\underbar{67}}: 180-224.

Banks, T. (1998b), ``The State of Matrix Theory'', {\underbar{Nuclear Physics
B -- Proceedings Supplements}} {\underbar{68}}: 261-267.

Bell, J. S. (1990), ``Against Measurement'', {\underbar{Physics World}}
{\underbar{August}}: 33-40.

Belot, G. and Earman, J. (1999), ``Pre-Socratic Quantum Gravity'',
in this volume.

Beltrametti, E. G. and Cassinelli, G. (1981), {\underbar{The Logic of Quantum
Mechanics}}. Reading, Massachusetts: Addison-Wesley.

Bohr, N. (1935), ``Can Quantum-Mechanical Description of Physical Reality
be Considered Complete?'', {\underbar{Physical}} {\underbar{Review}}
{\underbar{48}}: 696-702.

Bondi, H. (1952), ``Relativity and Indeterminacy'',
{\underbar{Nature}} {\underbar{169}}: 660-660.

Busch, P., Lahti, P.J., and Mittelstaedt, P. (1991),
{\underbar{The Quantum Theory of  Measurement}}. Berlin: Springer-Verlag.

Butts, D.A. and Rokhsar, D.S. (1999), ``Predicted Signatures of Rotating
Bose-Einstein Condensates'', {\underbar{Nature}} {\underbar{397}}: 327-329.

Castellano, M.G., Leoni, R., Torrioli, G., Carelli, P., and Cosmelli, C.
(1996), ``Development and Test of Josephson Devices for an Experiment of
Macroscopic Quantum Coherence'', {\underbar{IL Nuovo Cimento D}}
{\underbar{19}}: 1423-1428.

Choquet, G. and Meyer, P. (1963), ``Existence et Unicit\'e des
Repr\'esentations Int\'egrales dans les Convexes Compacts Quelconques'',
{\underbar{Annales Institut Fourier, Universite de Grenoble (France)}}
{\underbar{13}}: 139-154.

Christian, J. (1994),
``On Definite Events in a Generally Covariant Quantum World'',
Oxford University preprint.

Christian, J. (1996), ``The Plight of `I am' '',
{\underbar{Metaphysical Review}} {\underbar{3}}: 1-4
[also at quant-ph/9702012].  

Christian, J. (1997), ``Exactly Soluble Sector of Quantum Gravity'',
{\underbar{Physical Review D}} {\underbar{56}}: 4844-4877.

Christian, J. (1999a), ``Potentiality, Entanglement and
Passion-at-a-distance'', e-print quant-ph/9901008, to appear in
{\underbar{Studies in History and Philosophy of Modern Physics}.

Christian, J. (1999b), ``Evenhanded Quantum Gravity vs.
the World as a Hologram'', in preparation.

Cirac, J.I., Lewenstein, M., Molmer, K., and Zoller, P. (1998), ``Quantum
Superposition States of Bose-Einstein Condensates'',
{\underbar{Physical Review A}} {\underbar{57}}: 1208-1218.

Ciufolini, I., Pavlis, E., Chieppa, F., Fernandes-Vieira, E., and
P\'erez-Mercader, J. (1998), ``Test of General Relativity and Measurement
of the Lense-Thirring Effect with Two Earth Satellites'',
{\underbar{Science}} {\underbar{279}}: 2100-2103.

d'Espagnat, B. (1976),
{\underbar{Conceptual Foundations of Quantum Mechanics}}.
Reading, Massachusetts: Benjamin.

Di\'osi, L. (1984), ``Gravitation and Quantum-Mechanical Localization
of Macro-Objects'',
{\underbar{Physics}} {\underbar{letters A}} {\underbar{105}}: 199-202.

Di\'osi, L. (1987), ``A Universal Master Equation for the Gravitational
Violation of Quantum Mechanics'',
{\underbar{Physics}} {\underbar{letters A}} {\underbar{120}}: 377-381.

Di\'osi, L. (1989), ``Models for Universal Reduction of Macroscopic Quantum
Fluctuations'', {\underbar{Physical Review A}} {\underbar{40}}: 1165-1174.

Dixon, W. G. (1975), ``On the Uniqueness of the Newtonian Theory as a
Geometric Theory of Gravitation'',
{\underbar{Communications in Mathematical Physics}} {\underbar{45}}: 167-182.

Dum, R., Cirac, J.I., Lewenstein, M., and Zoller, P. (1998),
``Creation of Dark Solitons and Vortices in Bose-Einstein Condensates'',
{\underbar{Physical Review Letters}} {\underbar{80}}: 2972-2975.

Eddington, A. S. (1929),
{\underbar{The Nature of the Physical World}}. Cambridge, England: Cambridge
University Press.

Ehlers, J. (1981), ``\"Uber den Newtonschen Grenzwert der Einsteinschen
Gravitationstheorie'', in J. Nitsch, J. Pfarr, and E. W. Stachow (eds.),
{\underbar{Grundlagenprobleme der modernen Physik}}. Mannheim:
Bibliographisches Institut, pp. 65-84.

Ehlers, J. (1986), ``On Limit Relations Between, and Approximative Explanations
of, Physical Theories'', in R. Barcan Marcus, G. J. W. Dorn, and P. Weingartner
(eds.), {\underbar{Logic, Methodology and Philosophy of Science VII}}.
Amsterdam: North-Holland, pp. 387-403.

Ehlers, J. (1991), ``The Newtonian Limit of General Relativity'',
in G. Ferrarese (ed.), {\underbar{Classical Mechanics and}}
{\underbar{Relativity: Relationship and Consistency}}.
Naples: Bibliopolis, pp. 95-106.

Ehlers, J. (1997), ``Examples of Newtonian Limits of Relativistic Spacetimes'',
{\underbar{Classical and Quantum Gravity}} {\underbar{14}}: A119-A126.

Einstein, A. (1914), ``Die Formale Grundlage der allgemeinen
Relativit\"atstheorie'', {\underbar{K\"oniglich Preussische Akademie}}
{\underbar{der
Wissenschaften (Berlin)}} {\underbar{Sitzungsberichte}}: 1030-1085.

Einstein, A. (1994), {\underbar{Relativity, The Special and the General
Theory}}. London: Routledge.

Einstein, A., Podolsky, B., and Rosen, N. (1935), ``Can Quantum-Mechanical
Description of Physical Reality be Considered Complete?'',
{\underbar{Physical Review}} {\underbar{47}}: 777-780.

Ellis, J., Mohanty, S., and Nanopoulos, D.V. (1989), ``Quantum Gravity and
the Collapse of the Wavefunction'',
{\underbar{Physics Letters B}} {\underbar{221}}: 113-119.

Eppley, K. and Hannah, E. (1977), ``The Necessity of Quantizing the
Gravitational Field'', {\underbar{Foundations of Physics}} {\underbar{7}}:
51-68.

Feynman, R. (1957), a talk
given at the Chapel Hill Conference, Chapel Hill, North
Carolina, January 1957. [A report of the proceedings of the conference,
including discussions, can be obtained from Wright Air Development Center,
Air Research and Development Command, United States Air Force,
Wright-Patterson Air Force Base, Ohio, USA: WADC Technical Report 57-216.]

Feynman, R. (1995), {\underbar{Feynman Lectures on Gravitation}} (edited by
B. Hatfield). Reading, Massachusetts: Addison-Wesley 
[The section 1.4 from which the
quotation is taken was first presented by Feynman at the 1957 Chapel Hill
Conference (ibid), where, not surprisingly, it provoked an intense discussion].

Fivel, D. I. (1997),
``An Indication From the Magnitude of CP Violations That
Gravitation is a Possible Cause of Wave-Function Collapse'',
e-print quant-ph/9710042.

Frenkel, A. (1997), ``The Model of F. K\'arolyh\'azy and the Desiderata of
A. Shimony for a Modified Quantum Dynamics'', 
in R. S. Cohen, M. Horn, and J. Stachel (eds.),
{\underbar{Experimental Metaphysics: Quantum Mechanical}} {\underbar{Studies
for Abner Shimony, Volume One}}. Dordrecht: Kluwer Academic, pp. 39-59.

Ghirardi, G. C., Grassi, R., and Rimini, A. (1990),
``Continuous-Spontaneous-Reduction Model Involving Gravity'',
{\underbar{Physical Review A}} {\underbar{42}}: 1057-1064.

Haag, R. (1990), ``Fundamental Irreversibility and the Concept of Events'',
{\underbar{Communications in Mathematical}}
{\underbar{Physics}} {\underbar{132}}: 245-251.

Haag, R. (1992), {\underbar{Local Quantum Physics}}. Berlin: Springer-Verlag.

Hardy, L. (1998), ``An Experiment to Observe Gravitationally Induced
Collapse'', Oxford University preprint.

Hawking, S. W. and Ellis, G. F. R. (1973), {\underbar{The Large Scale Structure
of Space-time}}. Cambridge, England: Cambridge University Press.

Heisenberg, W. (1958),
{\underbar{Physics and Philosophy}}. New York: Harper and Row [see also
Aristotle, {\underbar{Metaphysics}}, Book IX (or Theta)].

Isham, C. J. (1993), ``Canonical Quantum Gravity and the Problem of Time'',
in L. A. Ibort and M. A. Rodriguez (eds.),
{\underbar{Integrable Systems, Quantum Groups, and Quantum Field
Theories}}. London: Kluwer Academic, pp. 157-288.

Isham, C. J. (1994), ``Prima Facie Questions in Quantum Gravity'', in J. Ehlers
and H. Friedrich (eds.), {\underbar{Canonical}}
{\underbar{Gravity: From Classical to
Quantum}}. Berlin: Springer-Verlag, pp. 1--21.

Isham, C. J. (1997), ``Structural Issues in Quantum Gravity'',
in E. Sorace, G. Longhi, L. Lusanna, and M. Francaviglia (eds.),
{\underbar{General Relativity and Gravitation: GR14}}.
Singapore: World Scientific, pp. 167-209.

Jauch, J. M. (1968), {\underbar{Foundations of Quantum Mechanics}}.
Reading, Massachusetts: Addison-Wesley.

Jones, K. R. W. (1995), ``Newtonian Quantum Gravity'',
{\underbar{Australian Journal of Physics}} {\underbar{48}}: 1055-1081.

Jordan, P. and Klein, O. (1927), ``Zum Mehrk\"orperproblem der
Quantentheorie'',
{\underbar{Zeitschrift f\"ur Physik}} {\underbar{45}}: 751-765.

K\'arolyh\'azy, F. (1966), ``Gravitation and Quantum Mechanics of Macroscopic
Bodies'', {\underbar{Nuovo Cimento A}} {\underbar{42}}:
390-402.

K\'arolyh\'azy, F., Frenkel, A., and Luk\'acs, B. (1986), ``On the Possible
Role of Gravity on the Reduction of the Wavefunction'', in R. Penrose and
C. J. Isham (eds.), {\underbar{Quantum Concepts in Space and Time}}.
Oxford, England: Clarendon Press, pp. 109-128.

Kay, B. S. (1998), ``Decoherence of Macroscopic Closed Systems within
Newtonian Quantum Gravity'', to appear in {\underbar{Classical and Quantum
Gravity}}, e-print hep-th/9810077.

Kent, A. (1990), ``Against Many-Worlds Interpretations'',
{\underbar{International Journal of Modern Physics A}}
{\underbar{5}}: 1745-1762 [A Forward has
been added to the updated e-print version located at gr-qc/9703089].

Kent, A. (1997), ``Consistent Sets Yield Contrary Inferences in Quantum
Theory'', {\underbar{Physical Review Letters}} {\underbar{78}}: 2874-2877.

Kibble, T. W. B. (1981), ``Is Semiclassical Theory of Gravity Viable?'',
in C. J. Isham, R. Penrose, and D. W. Sciama (eds.),
{\underbar{Quantum Gravity 2: a Second Oxford Symposium}}. Oxford, England:
Oxford University Press, pp. 63-80.

Komar, A. B. (1969), ``Qualitative Features of Quantized Gravitation'',
{\underbar{International Journal of Theoretical}}\n
{\underbar{Physics}} {\underbar{2}}: 157-160.

Kretschmann, E. (1917), ``\"Uber die prinzipielle Bestimmbarkeit der
berechtigten Bezugssysteme beliebiger Relativist\"atstheorien'',
{\underbar{Annalen der Physik (Leipzig)}} {\underbar{53}}: 575-614.

Kucha\v r, K. (1980), ``Gravitation, Geometry, and Nonrelativistic Quantum
Theory'', {\underbar{Physical Review D}} {\underbar{22}}: 1285-1299.

Kucha\v r, K. (1991), ``The Problem of Time in Canonical Quantization of
Relativistic Systems'', in A. Ashtekar and J. Stachel (eds.),
{\underbar{Conceptual Problems of Quantum Gravity}}. Boston:
Birkh\"auser, pp. 141-168.

Kucha\v r, K. (1992), ``Time and Interpretations of Quantum Gravity'',
in G. Kunstatter, D. Vincent, and J. Williams (eds.),
{\underbar{Proceedings of the 4th Canadian Conference on General
Relativity and Relativistic Astrophysics}}.\n
Singapore: World Scientific, pp. 211-314.

K\"unzle, H. P. (1972), ``Galilei and Lorentz Structures on Space-time:
Comparison of the Corresponding Geometry and Physics'',
{\underbar{Annales Institut Henri Poincar\'e}} {\underbar{A 17}}: 337-362.

K\"unzle, H. P. (1976), ``Covariant Newtonian Limit of Lorentz Space-Time'',
{\underbar{General Relativity and Gravitation}} {\underbar{7}}: 445-457.

Leggett, A. J. (1980), ``Macroscopic Quantum Systems and the Quantum Theory of
Measurement'', {\underbar{Supplement of}}
{\underbar{the Progress of Theoretical Physics}}
{\underbar{69}}: 80-100.

Leggett, A. J. (1984), ``Schr\"odinger's Cat and her Laboratory Cousins'',
{\underbar{Contemporary Physics}} {\underbar{25}}: 583-598.

Leggett, A. J. (1998), ``Macroscopic Realism: What Is It, and What Do We Know
about It from Experiment?'', in R. A. Healey and G. Hellman (eds.),
{\underbar{Minnesota Studies in the Philosophy of
Science, Volume XVII,}}\n
{\underbar{Quantum Measurement: Beyond Paradox}}. Minneapolis:
University of Minnesota Press, pp. 1-22.

Leggett, A. J. and Garg A. (1985), ``Quantum Mechanics versus Macroscopic
Realism: Is the Flux There when Nobody Looks?'',
{\underbar{Physical Review Letters}} {\underbar{54}}: 857-860.

Lense, J. and Thirring, H. (1918), ``\"Uber den Einfluss der Eigenrotation
der Zentralk\"orper auf die Bewegung der Planeten und Monde nach der
Einsteinschen Gravitationstheorie'', {\underbar{Physikalische Zeitschrift
(Germany)}} {\underbar{19}}:
156-163 [English translation by Mashhoon, B., Hehl, F. W., and Theiss, D. S.
(1984), ``On the Gravitational Effects of Rotating Masses -- The Thirring-Lense
Papers'', {\underbar{General Relativity and Gravitation}} {\underbar{16}}:
711-725].

Magnon, A. (1997), {\underbar{Arrow of Time and Reality: In Search of a
Conciliation}}. Singapore: World Scientific.

Misner, C. W., Thorne, K. S., and Wheeler, J. A. (1973),
{\underbar{Gravitation}}. New York: W. H. Freeman and Company.

Moroz, I. M., Penrose, R., and Tod, P. (1998),
``Spherically-symmetric Solutions of the Schr\"odinger-Newton Equations'',
{\underbar{Classical and Quantum Gravity}} {\underbar{15}}: 2733-2742.

Nakahara, M. (1990),
{\underbar{Geometry, Topology and Physics}}. Bristol: Adam Hilger.

Norton, J. D. (1993), ``General Covariance and the Foundations of General
Relativity: Eight Decades of Dispute'', {\underbar{Reports on Progress in
Physics}} {\underbar{56}}: 791-858.

Pearle, P. (1993), ``Ways to Describe Dynamical Statevector Reduction'',
{\underbar{Physical Review A}} {\underbar{48}}: 913-923.

Pearle, P. and Squires, E. J. (1996), ``Gravity, Energy Conservation, and
Parameter Values in Collapse Models'',
{\underbar{Foundations of Physics}} {\underbar{26}}: 291-305.

Penrose, R. (1979), ``Singularities and Time-Asymmetry'', in S. W. Hawking
and W. Israel (eds.), {\underbar{General}}
{\underbar{Relativity:}} {\underbar{an Einstein Centenary}}.
Cambridge, England: Cambridge University Press, pp. 581-638.

Penrose, R. (1981), ``Time-Asymmetry and Quantum Gravity'', in
C. J. Isham, R. Penrose, and D. W. Sciama (eds.),
{\underbar{Quantum Gravity 2: a Second Oxford Symposium}}. Oxford, England:
Oxford University Press, pp. 244-272.

Penrose, R. (1984), ``Donaldson's Moduli Space: a `Model' for Quantum
Gravity?'', in S. M. Christensen (ed.),
{\underbar{Quantum theory of Gravity}}. Bristol: Adam Hilger, pp. 295-298. 

Penrose, R. (1986), ``Gravity and State Vector Reduction'',
in R. Penrose and C. J. Isham (eds.),
{\underbar{Quantum Concepts}} {\underbar{in Space and Time}}. Oxford, England:
Clarendon Press, pp. 129-146.

Penrose, R. (1987), ``Newton, Quantum Theory and Reality'',
in S. W. Hawking and W. Israel (eds.),
{\underbar{300 Years of}}\n {\underbar{Gravity}}. Cambridge, England:
Cambridge University Press, pp. 17-49.

Penrose, R. (1989),
{\underbar{The Emperor's New Mind: Concerning Computers, Minds and the Laws of
Physics}}. Oxford, England: Oxford University Press.

Penrose, R. (1993), ``Gravity and Quantum Mechanics'',
in R. J. Gleiser, C. N. Kozameh and O. M. Moreschi (eds.),
{\underbar{General Relativity and Gravitation: GR13}}. Bristol: Institute of
Physics Publishing, pp. 179-189.

Penrose, R. (1994a), ``Non-Locality and Objectivity in Quantum State
Reduction'', in J. S. Anandan and J. L. Safko (eds.),
{\underbar{Fundamental Aspects of Quantum Theory}}. Singapore: World
Scientific, pp. 238-246.

Penrose, R. (1994b),
{\underbar{Shadows of the Mind: a Search for the Missing Science of
Consciousness}}. Oxford, England: Oxford University Press.

Penrose, R. (1996), ``On Gravity's Role in Quantum State Reduction'',
{\underbar{General Relativity and Gravitation}} {\underbar{28}}: 581-600.
Reprinted in this Volume.

Penrose, R. (1997), {\underbar{The Large, the Small and the Human Mind}}.
Cambridge, England: Cambridge University Press.

Penrose, R. (1998),
``Quantum Computation, Entanglement, and State Reduction'',
{\underbar{Philosophical Transactions of}} {\underbar{the Royal Society,
London}} {\underbar{A 356}}: 1927-1939.

Percival, I. C. (1995), ``Quantum Spacetime Fluctuations and Primary State
Diffusion'', {\underbar{Proceedings of the Royal}}
{\underbar{Society, London}} {\underbar{A 451}}: 503-513.

Polchinski, J. (1998), {\underbar{String Theory: Superstring Theory and
Beyond}}. Cambridge, England: Cambridge University Press.

Primas, H. (1983),
{\underbar{Chemistry, Quantum Mechanics and Reductionism: Perspectives in
Theoretical Chemistry}},\n
Second Corrected Edition. Berlin: Springer-Verlag.

Reichenbach, H. (1956), {\underbar{The Direction of Time}}. Berkeley:
University of California Press.

Rosen, G. (1972), ``Galilean Invariance and the General Covariance of
Nonrelativistic Laws'', {\underbar{American Journal of}} {\underbar{Physics}}
{\underbar{40}}: 683-687.

Rovelli, C. (1991), ``What is Observable in Classical and Quantum Gravity'',
{\underbar{Classical and Quantum Gravity}} {\underbar{8}}: 297-316. 

Rovelli, C. (1998),
``Strings, Loops and Others: a Critical Survey of the Present
Approaches to Quantum Gravity'', to appear in the proceedings of
{\underbar{General Relativity and Gravitation: GR15}},
Pune, India, 1997, e-print\n
gr-qc/9803024.

Saunders, S. (1996), ``Time, Quantum Mechanics, and Tense'',
{\underbar{Synthese}} {\underbar{107}}: 19-53.

Schr\"odinger, E. (1935), ``Die gegenw\"artige Situation in der
Quantenmechanik'', {\underbar{Naturwissenschaften}} {\underbar{23}}:
807-812; 823-828; 844-849 [English translation and commentary in Wheeler and
Zurek (1983)].

Schweber, S. S. (1961),
{\underbar{An Introduction to Relativistic Quantum Field Theory}}.
Evanston, Illinois: Row, Peterson and Company.

Sen, A. (1998), ``Developments in Superstring Theory'', e-print
hep-ph/9810356.

Shimony, A. (1963), ``Role of the Observer in Quantum Theory'',
{\underbar{American Journal of Physics}} {\underbar{31}}: 755-773.

Shimony, A. (1978), ``Metaphysical Problems in the Foundations of Quantum
Mechanics'', {\underbar{International}}\n {\underbar{Philosophical Quarterly}}
{\underbar{18}}: 3-17.

Shimony, A. (1989), ``Conceptual Foundations of Quantum Mechanics'', in
P. Davies (ed.), {\underbar{The New Physics}}. Cambridge, England: Cambridge
University Press, pp. 373-395.

Shimony, A. (1993a),
{\underbar{Search for a Naturalistic World View, Vol. I}}. Cambridge, England:
Cambridge University Press.

Shimony, A. (1993b),
{\underbar{Search for a Naturalistic World View, Vol. II}}. Cambridge, England:
Cambridge University Press.

Shimony, A. (1998), ``Implications of Transience for Spacetime Structure'',
in S. A. Huggett, L. J. Mason, K. P. Tod, S. T. Tsou, and N. M. J.
Woodhouse (eds.),
{\underbar{The Geometric Universe: Science, Geometry, and the Work of}}
{\underbar{Roger Penrose}}.
Oxford, England: Oxford University Press, pp. 161-172.

Smolin, L. (1998), ``Towards a background-independent approach to M-theory'',
e-print hep-th/9808192.

Sorkin, R. D. (1997), ``Forks in the Road, on the Way to Quantum Gravity'',
{\underbar{International Journal of Theoretical}} {\underbar{Physics}}
{\underbar{36}}: 2759-2781.

Stachel, J. (1993), ``The Meaning of General Covariance: The Hole Story'',
in J. Earman, A. Janis, G. Massey and N. Rescher (eds.),
{\underbar{Philosophical Problems of the Internal and External
Worlds: Essays on the Philosophy of}} {\underbar{Adolf Gr\"unbaum}}.
Pittsburgh, PA: University of Pittsburgh Press, pp. 129-160.

Stachel, J. (1994), ``Changes in the Concepts of Space and Time Brought About
by Relativity'', in C. C. Gould and R. S. Cohen (eds.),
{\underbar{Artifacts, Representations and Social Practice}}.
Dordrecht: Kluwer Academic, pp. 141-162.

Tod, P.
and Moroz, I.M. (1998), ``An analytic approach to the Schr\"odinger-Newton
equations'', to appear in {\underbar{Nonlinearity}}.

von Neumann, J. (1955), {\underbar{Mathematical Foundations of Quantum
Mechanics}}. Princeton: Princeton University Press.

Wald, R. M. (1984), {\underbar{General Relativity}}. Chicago:
University of Chicago Press.

Weinberg, S. (1989), ``Testing Quantum Mechanics'',
{\underbar{Annals of Physics (New York)}} {\underbar{194}}: 336-386.

Wheeler, J. A. and Zurek, W. H. (eds.) (1983),
{\underbar{Quantum Theory and Measurement}}. Princeton, New Jersey:
Princeton University Press.

Whitehead, A. N. (1929), {\underbar{Process and Reality}}. London: Macmillan.

Whitrow, G. J. (1961),
{\underbar{The Natural Philosophy of Time}}. Edinburgh: Nelson.

Zeilicovici, D. (1986) ``A (Dis)Solution of McTaggart's Paradox'',
{\underbar{Ratio}} {\underbar{28}}: 175-195.

\end